\documentclass[%
 reprint,
 amsmath,amssymb,
 aps,
floatfix,
]{revtex4-2}

\usepackage{graphicx}
\usepackage{dcolumn}
\usepackage{bm}
\usepackage{xcolor}
\usepackage{ulem}
\usepackage[caption=false]{subfig}

\usepackage[hidelinks]{hyperref}

\begin{document}

\preprint{APS/123-QED}

\title{Primordial trispectrum from kSZ tomography}

\author{Neha Anil Kumar}
 \email{nanilku1@jhu.edu}
\author{Gabriela Sato-Polito}%
\author{Marc Kamionkowski}
\author{Selim C. Hotinli}
\affiliation{%
 William H.\ Miller III Department of Physics and Astronomy, Johns Hopkins University
 Baltimore, MD 21218, USA
}%



\begin{abstract}
The kinetic Sunyaev Zel’dovich effect is a secondary CMB temperature anisotropy that provides a powerful probe of the radial-velocity field of matter distributed across the Universe. This velocity field is reconstructed by combining high-resolution CMB measurements with galaxy survey data, and it provides an unbiased tracer of matter perturbations in the linear regime. In this paper, we show how this measurement can be used to probe primordial non-Gaussianity of the local type, particularly focusing on the trispectrum amplitude $\tau_{\rm NL}$, as may arise in a simple two-field inflation model that we provide by way of illustration. Cross-correlating the velocity-field-derived matter distribution with the biased large-scale galaxy density field allows one to measure the scale-dependent bias factor with sample variance cancellation. We forecast that a configuration corresponding to CMB-S4 and VRO results in a sensitivity of $\sigma_{f_{\rm NL}} \approx 0.59$ and $\sigma_{\tau_{\rm NL}} \approx 1.5$. These forecasts predict improvement factors of 10 and 195 for $\sigma_{f_{\rm NL}}$ and $\sigma_{\tau_{\rm NL}}$, respectively, over the sensitivity using VRO data alone, without internal sample variance cancellation. Similarly, for a configuration corresponding to DESI and SO, we forecast a sensitivity of $\sigma_{f_{\rm NL}} \approx 3.1$ and $\sigma_{\tau_{\rm NL}} \approx 69$, with improvement factors of 2 and 5, respectively, over the use of the DESI data-set in isolation. We find that a high galaxy number density and large survey volume considerably improve our ability to probe the amplitude of the primordial trispectrum for the multi-field model considered.
\end{abstract}

\maketitle


\section{\label{sec:Introduction}Introduction}

Detecting and constraining characteristics of the primordial Universe to understand the origin of structure is one of the primary goals of many upcoming large-scale structure surveys and CMB experiments \cite{Ade:2018sbj, Abitbol:2019nhf,Abazajian:2016yjj, Abazajian:2022nyh, Sehgal:2019ewc, CMB-HD:2022bsz, Aghamousa:2016zmz, Schlegel:2019eqc}. The most widely accepted paradigm is that of inflation \cite{Guth:1980zm, Linde:1982bn, Albrecht:1982wi}, which addresses most of the problems of the original Big-Bang scenario and has a set of predictions that are compatible with  many current observations \cite{Hawking:1982cz, Guth:1982ec, Starobinsky:1982ee, Bardeen:1983qw}.  Although the general predictions of the inflationary model, such as a flat universe and largely scale-invariant set of initial fluctuations, have been confirmed by recent observations, the specific physical processes that govern this epoch are yet to be understood. 

Comparing the predictions of various inflationary models to astrophysical observations allows one to probe the physics of the ultra-high energy scales that are otherwise not directly accessible to experiments. Searching for imprints of primordial non-Gaussianity in the CMB spectrum or on the large-scale matter distribution are possible methods to effectively distinguish between various models of inflation and the number of degrees of freedom governing the epoch~\citep[e.g.,][]{Baumann:2011nk, Assassi:2012zq, Chen:2012ge, Pi:2012gf, Noumi:2012vr, Arkani-Hamed:2015bza, Gong:2013sma, Lee:2016vti, Kehagias:2017cym, Kumar:2017ecc, An:2017hlx, An:2017rwo, Baumann:2017jvh, Kumar:2018jxz, Goon:2018fyu, Anninos:2019nib, Kumar:2019ebj, Hook:2019zxa}. 

One simple, and widely studied class of such non-Gaussianity is the local-type or $f_{\rm NL}$-parametrization, in which one includes a quadratic term in the primordial potential $\Phi = \phi + f_{\rm NL}\phi^2$. In this model, both linear and quadratic terms in the potential originate from the same Gaussian field $\phi$, called the inflaton. The current best bound is $f_{\rm NL} = -0.9 \pm 5.1$, coming from the latest Planck satellite CMB analysis \cite{Planck:2019kim} with the growth factor normalized to one during matter domination.

Non-Gaussianity also naturally arises in models of inflation that involve more than one field. This could be due to a coupling term across the two fields \cite{Jeong:2012df} or the addition of a field with its own quadratic term \citep{Tseliakhovich:2010kf}. This can enhance the inflaton four-point function (or trispectrum) while not affecting the more widely considered three-point function (or bispectrum) \cite{Chen:2009zp,Suyama:2011qi}, making the primordial trispectrum a valuable signature of extra degrees of freedom in the early Universe. The amplitude of the primordial trispectrum has also been constrained by the Planck CMB data, with the most recent estimate being $\tau_{\rm NL} = (-5.8 \pm 6.5)\times 10^4$ \cite{Planck:2019kim}. 

Given Silk damping of the temperature fluctuations, there is not much room to significantly improve upon the $f_{\rm NL}$ measurements with CMB measurements alone. However, there are a few proposed methods to further constrain $\tau_{\rm NL}$ using its signature in the trispectrum of the 21-cm brightness temperature \cite{Cooray:2008eb}, the halo bias \cite{Yamauchi:2015mja,Sekiguchi:2018kqe}, and 3-point correlations between two-CMB-temperature and one-$\mu$-spectral-distortion fluctuations \cite{Bartolo:2015fqz}. The expected sensitivities for these proposals are $\tau_{\rm NL}\approx 50 - 100$.

In the case of local non-Gaussianity, we also expect to obtain constraints from the distribution of galaxies on large scales, relying only on the measurement of the galaxy power spectrum in the linear regime \cite{Munchmeyer:2018eey, Ferraro:2014jba}. This constraint can be obtained using the fact that a non-zero $f_{\rm NL}$ induces a scale-dependent bias factor \cite{Tseliakhovich:2010kf, Slosar:2008hx}, providing a unique signal that is not mimicked by changes in the other standard cosmological parameters. However, reaching the predicted multi-field threshold of $f_{\rm NL} \gtrsim 1$ remains difficult because of sample variance.

In this paper, we forecast the sensitivity of kSZ tomography to both $f_{\rm NL}$ and $\tau_{\rm NL}$, assuming that the primordial non-Gaussianity is induced by two different fields \cite{Tseliakhovich:2010kf}. The kinetic Sunyaev Zel’dovich (kSZ) effect is the secondary CMB temperature anisotropy induced by the peculiar velocity of interspersed free-electrons that scatter the CMB photons \cite{Sunyaev:1980nv, Zeldovich:1969ff, Zeldovich:1969sb, Sunyaev:1980vz, Sazonov:1999zp}. Cross-correlating high-resolution CMB maps with large-scale structure surveys will allow for the measurement of this kSZ contribution as a function of redshift, a technique termed kSZ tomography \cite{Zhang:2000wf, Ho:2009iw, Shao:2016uzt, Zhang:2010fa, Munshi:2015anr}. This cross-correlation can be used to re-construct the radial velocity field of free electrons in a 3-dimensional volume, from which the large scale matter distribution can be inferred. 

By comparing this kSZ-tomography-based matter perturbation amplitude with the amplitude of the galaxy power spectrum, one can obtain excellent constraints on the scale-dependent bias. Since the matter and galaxy distributions are determined independently, the bias can be measured on a mode-by-mode basis, thus circumventing the cosmic-variance limit that usually arises when inferring $f_{\rm NL}$ and $\tau_{\rm NL}$ from the galaxy distribution data-set in isolation \cite{Seljak:2008xr}.

We forecast the precision with which our model of primordial non-Gaussianity can be measured in two distinct scenarios: one in which only a single tracer from a galaxy survey is considered and another in which both the galaxy survey and the kSZ-reconstructed velocity field are jointly measured. The forecasts that include both the galaxy distribution and velocity field are based on the kSZ bispectrum formalism developed in Ref. \cite{Smith:2018bpn}, which accounts for photo-$z$ errors and the optical-depth degeneracy. We consider, in our forecasts, two baseline experimental configurations: `baseline 1' corresponding to the combination of Vera Rubin Observatory (VRO) \cite{LSSTScience:2009jmu} and CMB-S4 \cite{CMB-S4:2016ple} and `baseline 2' corresponding to Dark Energy Spectroscopic Instrument (DESI) \cite{DESI:2016fyo} and Simons Observatory (SO) \cite{Ade:2018sbj, Abitbol:2019nhf}. For these forecasts, we closely follow the method and experimental parameter values used in Ref. \cite{Munchmeyer:2018eey}, in which the same experimental configurations were used to forecast survey sensitivity to the single field $f_{\rm NL}$ model of inflation.

Our forecasts find that for the configuration of VRO and CMB-S4, $\sigma_{f_{\rm NL}} \approx 0.59$ and $\sigma_{\tau_{\rm NL}}\approx 1.5$, which corresponds to improvement factors of 10 and 195, respectively, over the use of VRO data alone. Similarly, for the configuration of the DESI and SO, we find that $\sigma_{f_{\rm NL}} \approx 3.1$ and $\sigma_{\tau_{\rm NL}} \approx 69$, with improvement factors of 2 and 5, respectively, compared to the forecasts made using DESI data in isolation. We find that our forecasts on the galaxy distribution data sets alone are compatible with the single-tracer results of  Ref. \cite{Ferraro:2014jba} when differences in our survey parameters are taken into account. 

Through the variation of experimental parameters we also determine that changes in redshift dispersion arising from photo-$z$ errors, as well as increases in CMB sensitivity and CMB resolution have a relatively minimal effect on our ability to measure the non-Gaussianity. In contrast, we find that a large survey volume, with well measured large-scale modes, and a high galaxy number density most prominently decrease the error with which both $f_{\rm NL}$ and $\tau_{\rm NL}$ can be measured.

Throughout this paper, we adopt the $\Lambda$CDM Cosmology as our fiducial model with the following parameters from \textit{Planck} 2018 \cite{Planck:2018vyg}: reduced Hubble constant $h = 0.674$, baryon and cold-dark-matter density parameters today $\Omega_b = 0.049$, and $\Omega_{\rm{cdm}} = 0.264$ respectively, spectral index $n_s = 0.965$ and amplitude of the primordial scalar power spectrum $A_s = 2.2\times10^{-9}$. In all our equations, we work under the convention $c=1$.

This paper is organised as follows. In Sec. \ref{sec:Theory}, we introduce our scale-dependent biasing model, derived using the peak-background-split methodology for the multi-field model of inflation presented in Ref. \cite{Tseliakhovich:2010kf}. We also explain how kSZ tomography can be used for sample variance cancellation.  In Sec. \ref{sec:Experiments}, we describe the experimental parameters in our forecast. The forecast set up is described in Sec. \ref{sec:FisherForecastSetUp}. Finally, the results of our analysis and our final set of forecasts are detailed in Sec. \ref{sec:Results}.

\section{\label{sec:Theory} Theory}
Before explaining in detail how kSZ tomography can be used in tandem with large-scale galaxy survey data, we first introduce the $\tau_{\rm NL}$ model of primordial non-Gaussianity and derive the relevant power spectra relations via the peak background split formalism. Further details on this model and its derivations can be found in Ref. \cite{Tseliakhovich:2010kf}. Furthermore, we address the possible extension of the results in this paper to another commonly considered, non-Gaussian model of inflation. Finally, we then also briefly address how the velocity field is reconstructed, given temperature-anisotropy and galaxy-distribution data.

\subsection{\label{subsec:PBS} Local non-Gaussianity in Peak Background Split Formalism}
In this paper, we consider the curvaton model for the primordial gravitational potential in which two different fields,  the inflaton and the curvaton, contribute to the curvature perturbation. The contribution from the inflaton is purely Gaussian while the perturbations of the curvaton field generate the non-Gaussianity. In this model, therefore, the primordial potential takes the following form:
\begin{eqnarray}
\Phi(\bm{x}) = \phi(\bm{x}) + \psi(\bm{x}) + f_{\rm NL}(1+\Pi)^2(\psi^2(\bm{x}) - \langle\psi^2\rangle).\ \ 
\label{eq:primPot}
\end{eqnarray}
Here, $\phi$ and $\psi$ are uncorrelated Gaussian random fields with power spectra that are proportional to each other, with proportionality constant $\Pi\equiv P_{\phi}/P_{\psi}$. For this model, the three and four point functions take the local form
\begin{eqnarray}
\xi_\phi^{(3)}(\bm{k}_1, \bm{k}_2, \bm{k}_3) = f_{\rm NL}[P_1P_2 + \text{5 perms.}] \\\nonumber + \mathcal{O}(f_{\rm NL}^3), \\
\xi_\phi^{(4)}(\bm{k}_1, \bm{k}_2, \bm{k}_3, \bm{k}_4) = 2\Big(\frac{5}{6}\Big)^2\tau_{\rm NL}[P_1P_2P_{13} \\\nonumber 
+\text{23 perms.}] + \mathcal{O}(\tau_{\rm NL}^2),
\end{eqnarray}
where we have defined $P_i \equiv P_{\Phi}(k_i)$, $P_{ij} = P_{\Phi}(|\bm{k_i} + \bm{k_j}|)$, and $\tau_{\rm NL} \equiv (6f_{\rm NL}/5)^2(1 + \Pi)$. Therefore, for $\Pi = 0$, this model reduces to the singly-parametrized $f_{\rm NL}$ model.

Large-scale halo bias is usually treated under the context of the peak background split formalism \cite{Cole:1989vx}, where one can split the density field into a long-wavelength piece $\delta_\ell$ and a short-wavelength piece $\delta_s$ as in
\begin{eqnarray}
\rho(\bm{x}) = \bar{\rho}(1 + \delta_\ell + \delta_s).
\end{eqnarray}

The local Lagrangian number density of halos $n(\bm{x})$,\ at position $\bm{x}$ is dependent on the local value of the long-wavelength perturbation $\delta_\ell$ as well as the local small scale power $\sigma_8^{\text{local}}(\bm{x})$. In the Gaussian case, since the small scale power is a constant, the Lagrangian bias is solely dependent on the variation in the halo number density as a function of the large-scale matter overdensity field \cite{Dalal:2007cu, Slosar:2008hx}.

When non-Gaussianity is present, the analysis under this formalism is complicated by the fact that the large- and small-scale density fluctuations are no longer independent. This becomes evident in the $\tau_{\rm NL}$ model when the long- and short-wavelength pieces of the Gaussian potential fluctuations are separated as follows:
\begin{align}
    \phi = \phi_\ell + \phi_s, && \psi = \psi_\ell + \psi_s.
\end{align}
Plugging this into Eq. \eqref{eq:primPot} will show that a few of the terms contain both short- and long-wavelength pieces. Therefore, this scenario needs more careful handling.

We start by establishing the Fourier-space relation between the primordial potential and matter overdensity field $\delta(\bm{k}, z) = \alpha(k, z)\Phi(\bm{k})$, where the form of the Poisson equation based operator $\alpha(k)$ is given by \cite{Slosar:2008hx}:
\begin{eqnarray}
\alpha(k, z) = \frac{2k^2T(k)G(z)}{3 \Omega_m H_0^2}.
\label{eq:alpha_def}
\end{eqnarray}
Here, $G(z)$ is the linear growth rate normalized such that $G(z) = 1/(1+z)$ during matter domination and $T(k)$ is the transfer function normalized to 1 at low $k$. Since this operator is usually defined in terms of its action in Fourier space, when applied to a real-space function such as $\phi(\bm{x})$, we use the convention
\begin{equation}
    \alpha \phi(\bm{x}) \equiv \int \frac{d^3\bm{k}}{(2\pi)^3}\alpha(k) e^{i\bm{k}\cdot \bm{x}} \int d^3\bm{y}\ \phi(\bm{y}) e^{-i\bm{k}\cdot \bm{y}}.
\end{equation}

With this relation in hand, the contributions from both the inflaton and the curvaton field, to the long-wavelength piece of the matter overdensity fluctuation, can be written as:
\begin{eqnarray}
\delta_\ell(\bm{x}) = \alpha[\phi_\ell(\bm{x}) + \psi_\ell(\bm{x})],
\end{eqnarray}
where the remaining terms are either much smaller ($f_{\rm NL}\psi_\ell^2$) or contain short-wavelength pieces. Similarly, within a region of given large-scale overdensity $\delta_\ell$ and potential $[\phi_\ell + \psi_\ell]$, the short-wavelength modes of the matter overdensity field are:
\begin{eqnarray}
\delta_s = \alpha[\phi_s + \psi_s(1 + 2f_{\rm NL} (1+\Pi)^2\psi_\ell)],
\end{eqnarray}
where the white-noise term [$f_{\rm NL}(1+\Pi)^2\psi_s^2$], that is spatially invariant when averaged over, has been disregarded, and the explicit $\bm{x}$-dependence of the terms has been dropped for ease of notation. 

Given the above split, it is evident that the mixing of the short- and long-wavelength pieces induces a scale dependence on the local small-scale power of the matter overdensity field:
\begin{equation}
    \begin{split}
        \sigma^2 &= \alpha^2\{\langle\phi_s^2\rangle + \langle\psi_s^2\rangle[1 + 2f_{\rm NL} (1+\Pi)^2\psi_\ell]^2\} \\
        &= \bar{\sigma}^2 [1 + 4f_{\rm NL} (1 + \Pi)\psi_\ell],
    \end{split}
\end{equation}
where $\bar{\sigma}^2 = \alpha^2\langle\psi^2_s\rangle(1+\Pi)$ and we have, once again, dropped any terms quadratic in $\psi_\ell$. The above expression indicates that when there exists primordial non-Gaussianity, the number density of halos varies not only with the large-scale matter overdensity modes but also with the local small-scale power.  This can be accounted for in the derivation of the Lagrangian halo bias as
\begin{eqnarray}
\delta_h \equiv \frac{\delta n_h}{n_h} = b_h\delta_\ell + \beta_f(1+\Pi)f_{\rm NL} \psi_\ell, 
\label{eq:deltaBiasBasic}
\end{eqnarray}
where
\begin{align}
    b_h \equiv \frac{\partial \ln{n_h}}{\partial \delta_\ell} && \text{and} && \beta_f \equiv 2\frac{\partial \ln{n_h}}{\partial \ln{\sigma}} = 2\delta_c(b_h - 1).
    \label{eq: beta_f}
\end{align}

Given the above form, it is straightforward to calculate the matter-halo ($P_{mh}$) and halo-halo ($P_{hh}$) power spectra. Using the fact that the inflaton and curvaton fields are uncorrelated, and that their power spectra are proportional to each other, one can derive
\begin{equation}
P_{mh}(k, z) = \Bigg[b_h + \beta_f \frac{f_{\rm NL}}{\alpha(k, z)}\Bigg]P_{mm}(k, z), 
\end{equation}
and 
\begin{equation}
    P_{hh}(k, z) = \Bigg[b_h^2 + 2b_h\beta_f\frac{f_{\rm NL}}{\alpha(k,z)} + \beta_f^2\frac{\big(\frac{5}{6}\big)^2\tau_{\rm NL}}{\alpha^2(k,z)}\Bigg]P_{mm}(k, z),
\end{equation}
where $P_{mm}(k, z)$ refers to the large-scale matter power spectrum, i.e., the Fourier space variance in our large-scale overdensity $\delta_\ell$. From this point on, since we will primarily be dealing with matter overdensities on linear scales, we will label our large-scale overdensity with $\delta_m$. On these linear scales, we will continue to use a subscript of $h$ when referring to the halo power spectra. 

The forecasts in this paper will, therefore, focus on calculating survey sensitivity to both $f_{\rm NL}$ and $\tau_{\rm NL}$ under the null hypothesis ($f_{\rm NL} = \tau_{\rm NL} = 0$), using both $P_{mh}$ and $P_{hh}$ as parametrized above. Although the parameter $\Pi$ more directly provides information on whether the primordial potential is defined by two different fields, we choose not to explore the parameter space in terms of $[f_{\rm NL}, \Pi]$. This is because $\Pi$ is defined as the ratio between two power spectra and can realistically be infinite in the absence of the curvaton field ($P_{\psi} = 0$).

Neverthless, we can still attempt to probe the degrees of freedom during inflation using the fact that the trispectrum amplitude satisfies $\tau_{\rm NL}\geq(6/5)^2 f_{\rm NL}^2$ for multi-field models. This relation can be used to define the parameter
\begin{equation}
r_{\rm NL}=(5/6)^2\tau_{\rm NL}-f_{\rm NL}^2\,,
\label{eq:rNL}
\end{equation}
which can deviate away from zero in the presence of additional degrees of freedom in the early Universe, depending on the value of $\Pi$. Hence, $\tau_{\rm NL}$ can be thought of as a probe of the extra degrees of freedom during inflation.

\subsection{\label{subsec:RSD} Redshift Space Distortions}
Redshift maps of galaxies distributed in a given survey volume are distorted by their peculiar velocities along the line of sight. When the bias relation is linear, the redshift-distorted halo overdensity is the sum of the biased matter overdensity in real space and a correction from the peculiar velocity of galaxies \cite{Kaiser:1987qv}
\begin{equation}
    \delta_{h, \text{RSD}}(\bm{x}) = b_1\delta_m(\bm{x}) + \frac{\partial}{\partial \bm{x}}\Big[\frac{\bm{u}(\bm{x})\cdot \hat{\bm{x}}}{aH}\Big],
\end{equation}
where $\bm{u}$ refers to the peculiar velocity of the galaxies and $\bm{x}$ refers to the position of the observed galaxy. To simplify the conversion to Fourier space, we use the late-time, linearized, continuity-equation-based relation between the peculiar-velocity field and matter-overdensity field, 
\begin{equation}
    \bm{u}(\bm{k}, z) = aHf \frac{i\bm{k}}{k^2}\delta_m(\bm{k}, z).
    \label{eq:velDeltam}
\end{equation}
Here, $f$ refers to the linear growth rate $d \ln{G}/ d \ln{a}$. With the above relation in hand, the Fourier transform of the redshift space linear bias relation simplifies to
\begin{equation}
    \delta_{h, \text{RSD}}(\bm{k}) = [b_h + f \mu_k^2]\delta_m(\bm{k})
\end{equation}
where $\mu_k$ is defined to be $\hat{\bm{e}}_z \cdot \hat{\bm{k}}$, the cosine of the angle between the line of sight and the wavevector $\hat{\bm{k}}$. 

It is straightforward to extend this derivation to the bias relation in Fourier space, for the $\tau_{\rm NL}$ model of primordial non-Gaussianity. The updated form of the halo overdensity is simply
\begin{equation}
    \delta_{h, \text{RSD}}(\bm{k})= [b_h + f \mu_k^2] \delta_m(\bm{k}) + \beta_f(1+\Pi)f_{\rm NL} \psi_\ell(\bm{k}),
\end{equation}
where the same correction term is added to the original form introduced in Eq.~\eqref{eq:deltaBiasBasic}. The power spectra can therefore be updated, under the effects of RSD, by replacing every instance of $b_h$ in our previously derived halo power spectra models with $b_{h, \text{RSD}} = [b_h + f \mu_k^2]$.

\subsection{Extension to the $g_{\rm NL}$ Model}
In this work, we primarily focus on $\tau_{\rm NL}$- and $f_{\rm NL}$- type non-Gaussianities that have a clear correspondence predicted in the case of single-field slow-roll inflation. However, the forecasts in this paper can be extended to another possible model, parametrized by $g_{\rm NL}$, in which the primordial potential takes the following form:
\begin{equation}
    \Phi(\bm{x}) = \phi(\bm{x}) + g_{\rm NL}[\phi^3(\bm{x}) - 3\langle\phi^2\rangle\phi(\bm{x})].
\end{equation}
For this single-field model, using the peak-background-split formalism from above, one can show that:
\begin{eqnarray}
    P_{mh}(k,z) = \Bigg[b_{h, \rm{RSD}} + \beta_g\frac{g_{\rm NL}}{\alpha(k,z)}\Bigg]P_{mm}(k,z), \\
    P_{hh}(k,z) = \Bigg[b_{h, \rm{RSD}} + \beta_g\frac{g_{\rm NL}}{\alpha(k,z)}\Bigg]^2P_{mm}(k,z)
\end{eqnarray}
where $\beta_g = 3 \partial \ln n_h / \partial f_{\rm NL}$. Since, the barrier crossing prediction for $\beta_g$ does not agree well with N-body simulations, previous forecasts on this model have used simulation-based fit-functions for $\beta_g$ that are independent of $f_{NL}$ under the null hypothesis (see, for example, Ref. \citep{Ferraro:2014jba}).

When compared with the single-field $f_{\rm NL}$-cosmology [where $\tau_{\rm NL} = (6/5 f_{\rm NL})^2$], one can see that the contribution of $g_{\rm NL}$ to the halo bias is the same as the contribution from $f_{\rm NL} = (\beta_f/ \beta_g) g_{\rm NL}$. Therefore, when only a single population of tracers is being considered to calculate $P_{hh}(k,z)$, the two parameters are indistinguishable. Since the forecasts in this paper are calculated using the cross-correlation of a single galaxy tracer with kSZ tomography, under the null hypothesis one can use these forecasts to obtain constraints on $g_{\rm NL}$ by expressing $\sigma_{g_{\rm NL}} = (\beta_f / \beta_g) \sigma_{f_{\rm NL}}$ \cite{Ferraro:2014jba}.

\subsection{\label{subsec:kSZ} The kSZ Effect}
The temperature fluctuation attributed to the kSZ in the $\hat{\bm{n}}$ direction in the sky is given by the integral \cite{Smith:2018bpn},
\begin{equation}
    T(\hat{\bm{n}}) = -T_{\text{CMB}}\sigma_T\int \frac{d\chi}{(1+z)} e^{-\tau(\chi)}n_e(\hat{\bm{n}}, \chi)\hat{\bm{n}}\cdot \bm{v},
\end{equation}
where $T_{\text{CMB}}$ is the average temperature of the CMB today, $\sigma_T$ is the Thomson Scattering cross-section, and $\tau$ is the optical depth to the scattering electron with velocity $\bm{v}$ at comoving distance $\chi$, and redshift $z$. The fluctuation is also dependent on the electron number density $n_e(\hat{\bm{n}}, \chi) = \bar{n}_e(\chi)(1 + \delta_e)$.

To use this anisotropy data and derive redshift dependent information one must cross-correlate the kSZ data set with a tracer of large-scale structure. Ref. \cite{Smith:2018bpn} shows that most of the varied approaches to this technique are equivalent to using a bispectrum of the form $\langle\delta\delta T\rangle$ to reconstruct the radial-velocity field. In the next few sections we summarize how this bispectrum is used to derive the expected form of the signal and noise. For a more detailed derivation of the results, see Ref. \cite{Smith:2018bpn}.

\subsubsection{\label{subsubsec:kSZBispec} Bispectrum Based Estimator}
According to Ref. \cite{Smith:2018bpn}, the statistic that carries the kSZ tomography signal is a 3-point function defined as, 
\begin{equation}
    \langle\delta_X(\bm{k})\delta_X(\bm{k}')T(\ell)\rangle = B(\bm{k}, \bm{k'}, \ell)(2\pi)^3\delta^3\Bigg(\bm{k} + \bm{k}' + \frac{\ell}{\chi_*}\Bigg),
\end{equation}
where $\delta_X$ refers to the overdensity of the tracer in consideration, and all terms marked with a $*$ refer to quantities evaluated at redshift $z_*$. It can be shown that the kSZ is dominant in the squeezed limit \cite{Smith:2018bpn}, in which the bispectrum takes the form:
\begin{equation}
    B(k_L, k_S, \ell, k_{Lr}) = - \frac{K_* k_{Lr}}{\chi_*^2}\frac{P_{Xv}(k_L)}{k_L}P_{Xe}(k_S),
    \label{eq: bispectrum}
\end{equation}
where $k_L$ refers to the long-wavelength mode, $k_{Lr}$ is its component along the line of sight, $k_S$ refers to the short-wavelength mode, and $P_{Xv}$ and $P_{Xe}$ refer to the cross-spectra of the tracer overdensity field with the velocity field, and the electron density perturbations, respectively. In the above equation we have also defined,
\begin{equation}
    K_* \equiv -T_{\text{CMB}}\sigma_T\bar{n}_{e,0}e^{-\tau(\chi_*)}(1+z_*)^2,
\end{equation}
where $\bar{n}_{e,0}$ is the mean electron density today.

\subsubsection{\label{subsubsec:Nvv} Velocity Reconstruction}
As shown in Ref. \cite{Smith:2018bpn}, a quadratic estimator for the long-wavelength velocity modes can be constructed by summing over the pairs [$\delta_X(\bm{k}_S)T(\bm{\ell})$] of short-wavelength modes in the galaxy and CMB maps. This method is equivalent to the optimal kSZ bispectrum estimator \cite{Smith:2018bpn}.

Given the form of $B(k_L, k_S, \ell, k_{Lr})$ in Eq. \eqref{eq: bispectrum}, the signal to noise ratio $S/N$ of the kSZ bispectrum in the squeezed limit is 
\begin{equation}
    \frac{S}{N} = V \int \frac{d^3\bm{k}_L}{(2\pi)^3}\frac{k_{Lr}^2}{k_L^2}\frac{P_{gv}(k_L)^2}{P_{gg}^{\text{tot}}(k_L)N_{v_r}(k_L)}, 
    \label{eq:vel_SNR}
\end{equation}
where $N_{v_r}$ is the noise associated with radial velocity reconstruction. This noise is modelled as
\begin{equation}
    N_{v_r}(k_L, \mu_L) = \frac{2\pi \chi_*^2}{K_*^2}\Bigg[\int dk_S \frac{k_SP_{ge}(k_S)^2}{P_{gg}^{\text{tot}}(k_S)\ C_{\ell=k_S\chi_*}^{\text{tot}}}\Bigg]^{-1}.
    \label{eq:NvvFullExpression}
\end{equation}
In the above two equations, we have explicitly used a subscript of $g$ to label our tracer $X$. Therefore, $P_{gg}(k_{S}, \mu_{S})$ refers to the small-scale galaxy-galaxy auto-power spectrum and $P_{ge}(k_{S}, \mu_{S})$ is the small-scale galaxy-electron power spectrum. Finally, $\mu_L$ refers to the angle of the large-scale mode with respect to the line of sight, i.e. $\mu_L = \hat{\bm{k}}_L\cdot \hat{\bm{n}}$. However, it is important to note that $\mu_S$ and $\mu_L$ are not independent of each other. The value of $\mu_S$ is completely determined by $k_L, \mu_L$ and $k_S$ since the line of sight components of the Fourier modes $k_L$ and $k_S$, are equal to each other. The total noise in our velocity reconstruction $N_{vv}$ is then $N_{vv} = \mu_{L}^{-2}N_{v_r}$.

Here and below, a subscript of `$g$' will be used to denote small-scale galaxy power spectra that appear in kSZ tomography, in contrast to the subscript $h$ that has so far been used to label large-scale halo power spectra. While on large scales, we will assume that a single galaxy occupies each halo, small scale galaxy power spectra will be calculated within the halo model including the halo occupation distribution (HOD) \cite{Leauthaud:2011rj, Leauthaud:2011zt}. The modelling assumptions and parameter values used to construct the small-scale spectra under this model can be found in Appendix \ref{appendix:HaloModel}. 

In our model of the velocity reconstruction noise, we also include the effect of photo-$z$ errors via a Gaussian kernel of the form
\begin{equation}
    W^2(k, \mu) = e^{-k^2\mu^2\sigma^2(z)/H^2(z)},
    \label{eq:gaussKernel}
\end{equation}
where $\sigma(z)$ is the redshift scattering of the galaxy survey in consideration. This induces a $\mu_S = \hat{\bm{k}}_S\cdot \hat{\bm{n}}$ dependence on the small-scale galaxy-galaxy and galaxy-electron spectra. Further details on the noise in the kSZ velocity reconstruction due to photo-$z$ errors can be found in Ref. \cite{Smith:2018bpn}.

Finally, based on the linear relation between matter overdensities and peculiar velocities [Eq.~\eqref{eq:velDeltam}], the noise in the reconstructed density perturbation field is
\begin{equation}
    N_{mm}^{\text{rec}}(k_L, \mu) = \frac{k_L^2}{(faH)_*^2}N_{vv}(k_L, \mu).
    \label{eq:VelReconNvv}
\end{equation}
It is important to note that the noise is proportional to the magnitude $k_L^2$. This implies that the reconstruction noise is lowest on largest scales, which corresponds to the regime where cosmic variance is a dominant noise source. Therefore, it is on these scales that we would expect this independent probe of large-scale structure to significantly contribute to sample variance cancellation.

\section{\label{sec:Experiments} Experiment Specifications}
The primary set of forecasts presented in this paper consider two next generation large-scale structure experiments, DESI and VRO. VRO is an example of a high number density galaxy survey with photometric redshifts. In contrast, DESI is a low number density survey with precise, spectroscopic redshifts. Our forecasts assume a cross-correlation of these data-sets with kSZ data from CMB-S4, as well as data from a configuration similar to that of SO, to test and display the effects of sample variance cancellation.

\subsection{\label{subsec:LSSExperiments} Large-Scale Structure Experiments}
For our forecasts on VRO, we use the specifications for the \textit{LSST Gold Sample} as prescribed in the LSST science book \cite{LSSTScience:2009jmu}. The galaxy number density for this data-set, per $\text{arcmin}^2$, is given by
\begin{equation}
    n(z) = n_{0}\frac{1}{2z_0}\Big(\frac{z}{z_0}\Big)^2\exp({-z/z_0}),
\end{equation}
with $z_0 = 0.3$ and $n_{0} = 40\ \text{arcmin}^2$. At $z = 1$, this corresponds to a galaxy number density of approximately $n_\text{gal} = 10^{-2} \ \text{Mpc}^{-3}$. The photometric redshift error for this survey is 
\begin{equation}
    \sigma_z = 0.03(1+z),
\end{equation}
 which corresponds to a redshift dispersion of 0.06 at $z = 1$. Finally, the bias for this sample is also specified to be
 \begin{equation}
     b(z) = 0.95/G(z),
 \end{equation}
with their growth factor normalized such that $G(z = 0) = 1$. This corresponds to a bias of 1.6 at a redshift of 1. For DESI, we make a single forecast assuming a galaxy number density of $n_{\text{gal}} = 10^{-4}\ \rm{Mpc}^{-3}$ with a Gaussian halo bias of 1.6 at redshift 1, in accordance with the specifications provided in the DESI white-paper \cite{DESI:2016fyo}. 

To ensure that the small-scale power spectra generated based on the HOD model are consistent with the specifications of the experiments in consideration, we use the following prescription. In the HOD model, the galaxy sample is specified by imposing a particular threshold stellar mass $m_{\star}^{\text{thresh}}$ of observable galaxies. Since the remaining parameters defining the galaxy distribution for $P_{gg}$ and $P_{ge}$ are dependent on this parameter, we match the value of $m_{\star}^{\text{thresh}}$ so that the total predicted galaxy number density matches number density expected for a given experiment. The details on these power spectra's dependencies on $n_\text{gal}$ and $m_{\star}^{\text{thresh}}$ can be found in Appendix \ref{appendix:HaloModel}.

\subsection{\label{subsec:CMBExperiments} CMB Experiments}
Most of our forecasts are based on the planned CMB-S4 experiment specifications. Although the exact instrument specifications are still pending, we assume an effective beam with full-width-half-maximum (FWHM) of 1.5 arcminutes and a sensitivity of $1.0 \ \mu \rm{K}-$arcmin, which is one of many possible configurations. The effects of atmospheric noise are not included since they are expected to be sub-dominant to the instrument and kSZ contributions at the relevant high multipoles of $\ell > 3000$. The final set of contributions to the CMB spectrum that enters Eq.~\eqref{eq:NvvFullExpression} can be written as
\begin{equation}
    C_{\ell}^{\text{tot}} = C_{\ell}^{TT} + C_{\ell}^{\text{kSZ-late-time}} + N_{\ell}.
    \label{eq:Cll_contributions}
\end{equation}
Here, $C_{\ell}^{TT}$ is the lensed CMB temperature power spectrum, $C_{\ell}^{\text{kSZ-late-time}}$  is the low redshift contribution to kSZ and finally $N_{\ell}$ is the instrumental noise power spectrum of the CMB map, which is modelled as
\begin{equation}
    N(\ell) = s^2\text{exp}\Bigg[\frac{\ell(\ell + 1)\theta_{\rm FWHM}^2}{8\ \text{ln}2}\Bigg],
    \label{eq:CMBNoise}
\end{equation}
where $s$ labels the sensitivity of the instrument and $\theta_{\rm FWHM}$ is the resolution. We also make a forecast for a configuration with noise and beam comparable to SO. To make this estimate we use a beam with a resulotion of 1.5 armcmin and an effective white noise level of $5.0\ \mu \rm{K}-$arcmin, matching the set up in Ref. \cite{Munchmeyer:2018eey}.

\section{\label{sec:FisherForecastSetUp} Forecast Setup}
In this section, we briefly describe the construction of the information matrix and the relevant systematics, focusing on the methodology used for forecasts on the cross-correlated data-sets. We then establish the models and parameter space over which the information matrix is constructed. 

For our forecast, the measured modes are the large-scale modes of $[v(\bm{k}), \delta_h(\bm{k})]$ where $\{v(\bm{k})\}$ are the kSZ velocity reconstruction modes and $\{\delta_h(\bm{k})\}$ are large-scale halo overdensity modes. The halo-overdensity modes are obtained from the  survey data set assuming that each halo is occupied by exactly one galaxy. Therefore, the signal and noise matrices are
\begin{eqnarray}
    \textbf{S}(k, \mu, z) = \begin{pmatrix}P_{vv} & P_{vh} \\ P_{vh} & P_{hh}\end{pmatrix},\\
    \textbf{N}(k, \mu, z) = \begin{pmatrix} N_{vv} & 0 \\ 0 & N_{hh}\end{pmatrix}.
\end{eqnarray}
The covariance matrix of our measured signal is the sum of the above two matrices,
\begin{equation}
    \textbf{C}(k, \mu, z) = \textbf{S}(k, \mu, z) + \textbf{N}(k, \mu, z).
\end{equation}
The information matrix at redshift bin $z_*$ is, therefore,
\begin{equation}
    \begin{split}
        F_{ab} &= \frac{V}{2}\int \frac{d^3k}{(2\pi)^3}\text{Tr}[\textbf{C}(\bm{k})_{,a}\textbf{C}(\bm{k})^{-1}\textbf{C}(\bm{k})_{,b}\textbf{C}(\bm{k})^{-1}] \\
        &= \frac{V}{2}\int \frac{k^2 dk\  d\mu}{(2\pi)^2}\text{Tr}[\textbf{C}(\bm{k})_{,a} \textbf{C}(\bm{k})^{-1} \textbf{C}(\bm{k})_{,b}\textbf{C}(\bm{k})^{-1}],
    \end{split}
    \label{eq:FisherIntegral}
\end{equation}
where we have accounted for the fact that the covariance-matrix elements are only dependent on $k$ and $\mu$, with the latter being induced by the kSZ based velocity reconstruction and the inclusion of photo-$z$ errors. We assume that the integral can be performed from a lower limit $k_\text{min} \equiv \pi/V^{1/3}$, restricted by the survey volume $V$, to an upper limit $k_\text{max} \approx 10^{-1} \ \text{Mpc}^{-1}$. 

The final models for $P_{hh}(k, \mu, z)$ and $P_{vh}(k, \mu, z)$, as they appear in the covariance matrix, are:
\begin{eqnarray}
    P_{vh}(k) = \frac{b_v f a H}{k}\Bigg[ b_{h,{\rm RSD}}+ \beta_f \frac{f_{\rm NL}}{\alpha(k)}\Bigg]P_{mm}(k) \\
    P_{hh}(k) = \Bigg[b_{h,{\rm RSD}}^2 +  2b_{h,{\rm RSD}}\beta_f\frac{f_{\rm NL}}{\alpha(k)} + \\\nonumber \beta_f^2\frac{\big(\frac{5}{6}\big)^2\tau_{\rm NL}}{\alpha^2(k)}\Bigg]P_{mm}(k),
     \label{eq:final_models_for_Fisher}
\end{eqnarray}
where $b_{h,{\rm RSD}} = [b_h + f\mu^2]$, and the explicit dependence of some terms on $z$ and $\mu$ have been dropped for ease of notation. These models are a direct result of the derivations in Sec. \ref{sec:Theory}.

To model the final signal term $P_{vv}(k,z)$, we use the relation between the velocity and matter power spectra introduced in Eq.~\eqref{eq:velDeltam},
\begin{equation}
    P_{vv}(k, z) = \Bigg(\frac{b_v f a H}{k}\Bigg)^2P_{mm}(k, z).
    \label{eq: Pvv model}
\end{equation}
Here, we have introduced a the optical-depth degeneracy parameter $b_v$. This parameter, with an expected value of 1.0, is introduced to account for the fact that kSZ data allows for the measurement of the product of the $P_{ge}$ and $P_{gv}$, which means that a constant factor of scale could be exchanged between the two while keeping the signal unchanged.

In summary, the measurement covariance matrix $\textbf{C}(k, \mu, z)$ is constructed based on the above three models for $P_{hh}(k, \mu, z)$, $P_{vh}(k, \mu, z)$ and $P_{vv}(k, z)$. This is used to construct a $4\times4$ information matrix over the parameter space spanned by $[b_h,\ b_v,\ f_{\rm NL},\ \tau_{\rm NL}]$ with the fiducial values set to [$1.6,\ 1.0,\ 0.0,\ 0.0$], respectively. We invert this matrix and marginalise over the parameters $b_h$ and $b_v$ to obtain error estimates for $f_{\rm NL}$ and $\tau_{\rm NL}$. We also experimented with marginalising over cosmological parameters, but found that these do not significantly change our error estimates.

It is important to note that the value of $\Pi$, under the curvaton model constraint $\tau_{\rm NL} = (6f_{\rm NL}/5)^2(1+\Pi)$, is not well-defined for the assumed null hypothesis, $f_{\rm NL} = \tau_{\rm NL} = 0$. Therefore, in our forecasts, we assume that the models presented in Eq.~\eqref{eq:final_models_for_Fisher} represent one possible parametrization of non-Gaussian power spectra under a multi-field model of inflation. That is, we vary the parameters $\tau_{\rm NL}$ and $f_{\rm NL}$ independently, around their fiducial values, to construct our information matrix and make our forecasts. These estimates are propagated to quote a constraint for the parameter $r_{\rm NL}$ introduced in Eq.~\eqref{eq:rNL}.

The noise spectrum for the velocity reconstruction term $N_{vv}$ is given by $\mu_{L}^{-2}N_{v_r}$, where the form of $N_{v_r}$ was introduced in Eq.~\eqref{eq:VelReconNvv}. For halos, we assume that the noise is given primarily by the galaxy shot noise along with photo-$z$ errors. Photo-$z$ errors can be implemented for halos by a convolution of the halo density field with a Gaussian kernel in the radial direction, the form of which was introduced in Eq.~\eqref{eq:gaussKernel}. The halo noise power spectrum is then
\begin{equation}
    N_{hh}(k, \mu) = \frac{1}{W^2(k, \mu)n_\text{gal}},
\end{equation}
where we have directly used the galaxy number density $n_\text{gal}$ based on our assumption that the galaxy distribution has a one-to-one correspondence with the distribution of halos, on large scales.

\section{\label{sec:Results}Forecast Results}

In this section we provide forecasts for different experimental configurations. In the first part we analyse two different baseline configurations and provide estimates based on our assumed specifications on the galaxy survey and CMB measurement instrumentation. We then consider one of these baselines and vary each of the esperimental parameters, in isolation, to display the effects of these variations on our ability to constrain $f_{\rm NL}$ and $\tau_{\rm NL}$.

\subsection{\label{subsec:BaselineForecasts} Baseline Forecasts}

\begin{figure*}
    \centering
    \includegraphics[width=\columnwidth]{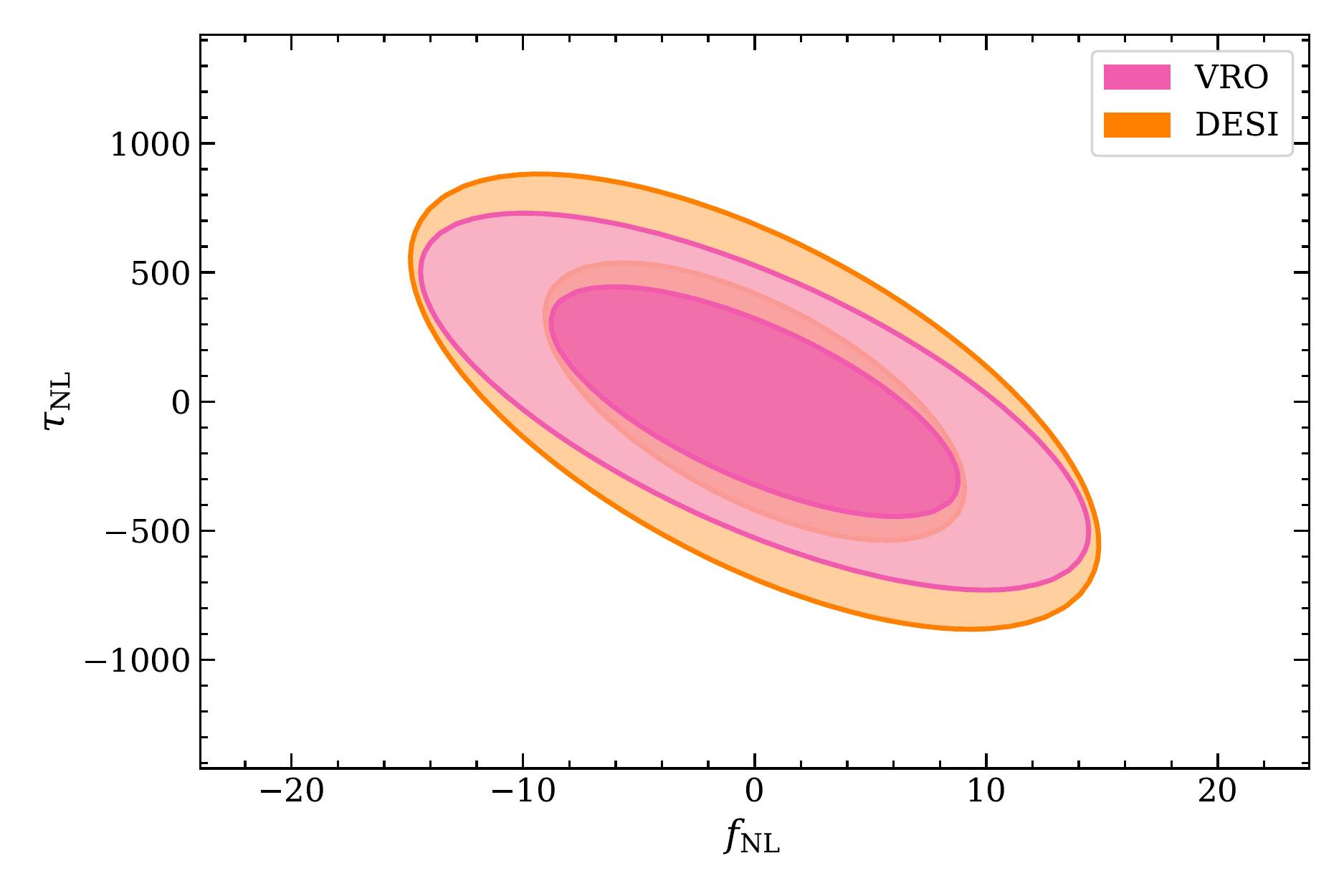}%
    \centering
    \includegraphics[width=\columnwidth]{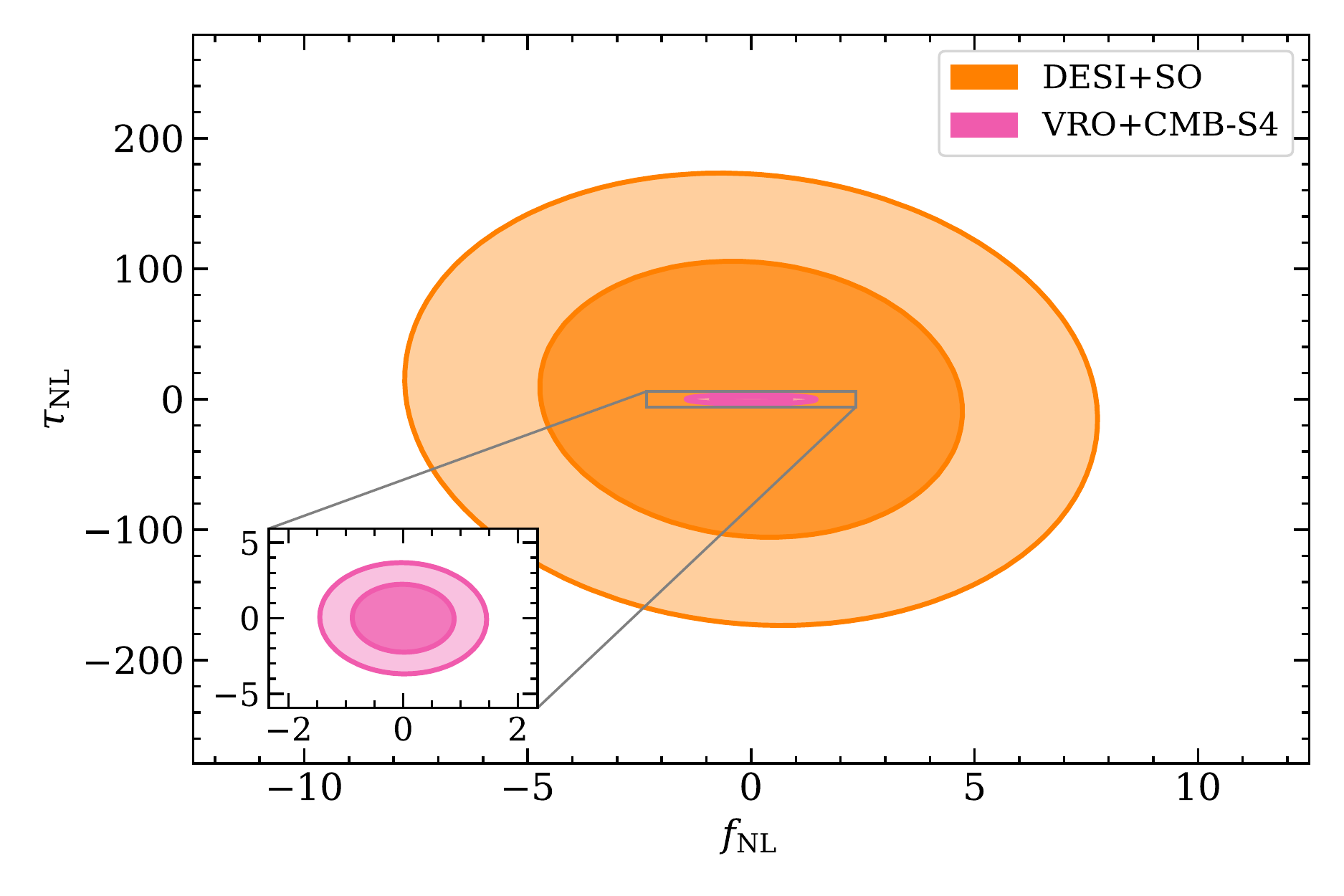}%
    \caption{Forecasted error ellipses on $f_{\rm NL}$ and $\tau_{\rm NL}$ at $68\%$ and $95\%$ confidence intervals, after marginalizing over $b_{h}$ and $b_v$. \textit{Left:} results when only galaxy survey data is considered. \textit{Right:} results when velocity reconstruction data is added to the analysis. Each color corresponds to one of the baselines defined in Table \ref{tab:baselineSpecs}.}%
    \label{fig:confidenceEllipses}%
\end{figure*} 

To establish the parameter dependencies of our forecast, we first display the results from our information matrix analysis for the two sets of baseline experiments described in Sec. \ref{sec:Experiments}. Their instrumental specifications have been summarized in Table \ref{tab:baselineSpecs}. Baseline 1 specifications were chosen to resemble the experimental configuration of VRO and CMB-S4. Similarly, baseline 2 corresponds to the combination of DESI and an SO-like CMB experiment. 

\begin{table}
\caption{\label{tab:baselineSpecs}%
Baseline configurations for the cross-correlated CMB and LSS experiments. Values for baseline 1 match the specifications of the VRO survey and CMB-S4. The values for baseline 2 are similar to those expected for DESI and SO. The survey volumes were kept the same across the two forecasts to emphasize the dependence of the forecasts on galaxy density and photo-$z$ errors.
}
\begin{ruledtabular}
\begin{tabular}{cccc}
& & \textrm{baseline 1} & \textrm{baseline 2}\\
\colrule
redshift & $z$ & 1.0 & 1.0\\
survey volume & $V$ & 100 $\ \text{Gpc}^{3}$ & 100$\ \text{Gpc}^3$\\
halo bias & $b_h$ & 1.6 & 1.6\\
galaxy density & $n_{\rm{gal}}$ & $10^{-2}\ \ \text{Mpc}^{-3}$ & $2\times10^{-4}\ \ \text{Mpc}^{-3}$\\
photo-$z$ error & $\sigma_z$ & 0.06 & - \\
CMB resolution & $\theta_{\rm FWHM}$ & 1.5 arcmin & 1.5 arcmin\\
CMB sensitivity & $s$ & 1 $\ \mu \rm{K}-$arcmin & 5 $\ \mu \rm{K}-$arcmin\\
\end{tabular}
\end{ruledtabular}
\end{table} 

Our forecasts on the aforementioned baseline configurations have been summarized in Table \ref{tab:baselineResults}.  This table also includes constraints on the parameter $r_{\rm NL}$, around a fiducial value of 0, corresponding to the assumed null hypothesis. To simplify our calculations, we assume a cubic geometry for the survey volume and for our kSZ formalism. Therefore, these forecasts do not include the effects of the time evolution of power spectra and biases on the light cone. For a complementary kSZ formalism using maps on the light-cone, see e.g., Refs.~\cite{Terrana:2016xvc, Deutsch:2017ybc, Cayuso:2021ljq}. 

For the baseline 1 experiments, the improvement factor in our ability to measure $f_{\rm NL}$ and $\tau_{\rm NL}$, arising from the cross-correlation with kSZ data, is 10 and 195, respectively. For baseline two, the improvement is 2 and 5 for the two parameters, respectively. In both cases, the improvement factor in the correlation-coefficient between the two parameters is much higher. This is explained by the fact that the cross-correlation of the two data sets allows for the inclusion of the $P_{vh}(k)$ signal, which offers an independent constraint for $f_{\rm NL}$. 

\begin{table}[b]
\renewcommand{\arraystretch}{1.5}
\caption{\label{tab:baselineResults}%
Information matrix based estimates on the errors across the two parameters, $f_{\rm NL}$ and $\tau_{\rm NL}$. The results of error propagation to $r_{\rm NL}$ are also included. To arrive at these estimates, a mean value of 0 was chosen for both $f_{\rm NL}$ and $\tau_{\rm NL}$, while the halo bias was defined for each experiment as shown in Table \ref{tab:baselineSpecs}.
}
\begin{ruledtabular}
\begin{tabular}{clcc}
& & \textrm{baseline 1} & \textrm{baseline 2}\\
\colrule
$f_{\rm NL}$ error  & $\sigma_{f_{\rm NL}}^\text{gal}$ & 5.8 & 6.0 \\
& $\sigma_{f_{\rm NL}}^\text{kSZ+gal}$ & $5.9\times10^{-1}$ & 3.1 \\
$\tau_{\rm NL}$ error & $\sigma_{\tau_{\rm NL}}^\text{gal}$ & $2.9 \times 10^{2\text{ }}$ & $3.6 \times 10^2$ \\
& $\sigma_{\tau_{\rm NL}}^{\text{kSZ+gal}}$ & $1.5$ & $6.9\times10^{1}$\\
$r_{\rm NL}$ error & $\sigma_{r_{\rm NL}}^\text{gal}$ & $2.0 \times 10^{2\text{ }}$ & $2.5 \times 10^2$ \\
& $\sigma_{r_{\rm NL}}^{\text{kSZ+gal}}$ & $1.0$ & $4.8\times 10^{1}$\\
\end{tabular}
\end{ruledtabular}
\end{table} 

\subsection{Experiment Parameter Variations}
In order to assess which experimental limitations have the greatest impact on measurements of primordial non-Gaussianity, we isolate the effects of certain experimental parameters on our ability to constrain $f_{\rm NL}$ and $\tau_{\rm NL}$ by varying each parameter in isolation. For the following forecasts, we assume the baseline 1 configuration, the specifics of which are provided in Table \ref{tab:baselineResults}.

\begin{figure*}
    \centering
    \includegraphics[width=\columnwidth]{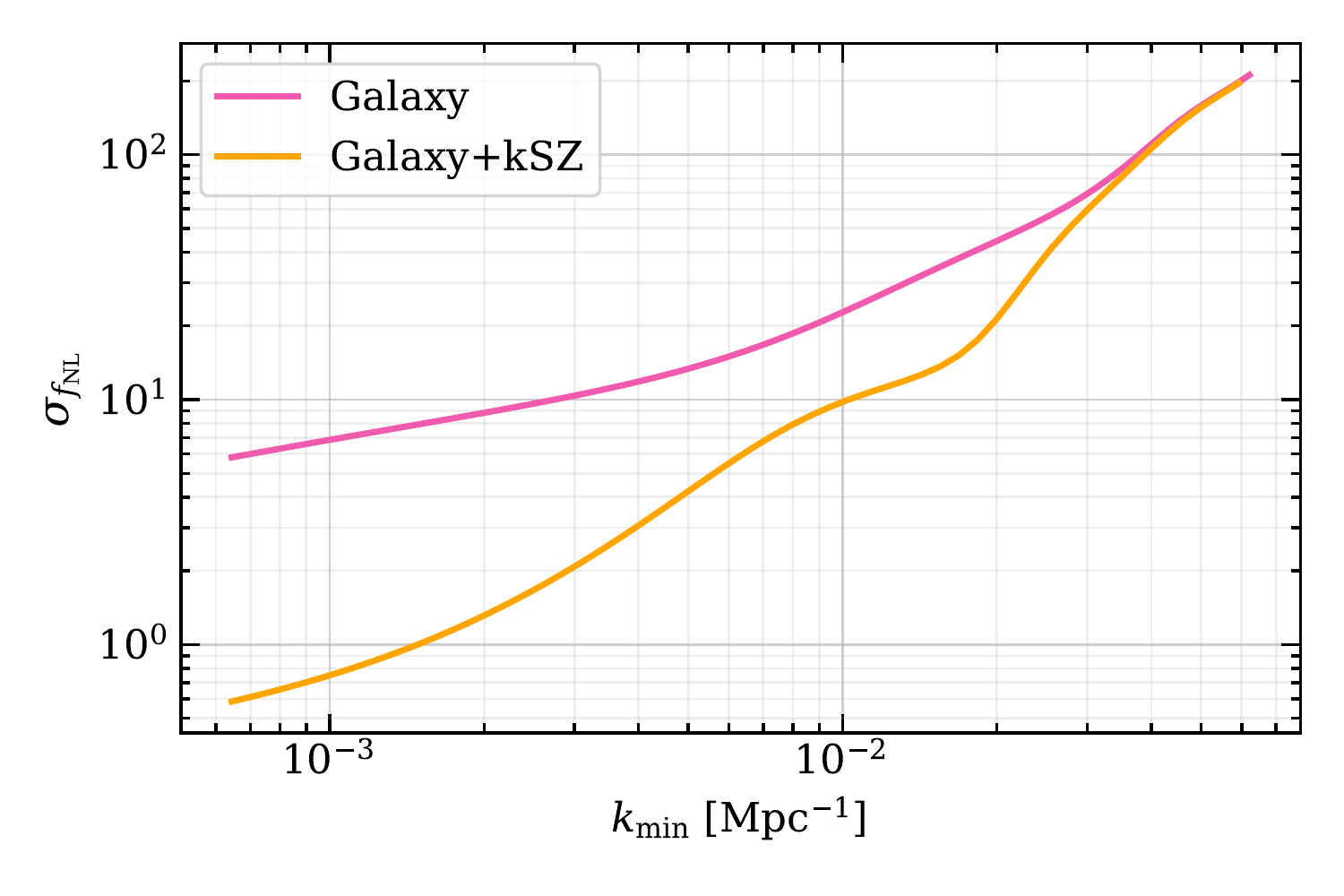}%
    \centering
    \includegraphics[width=\columnwidth]{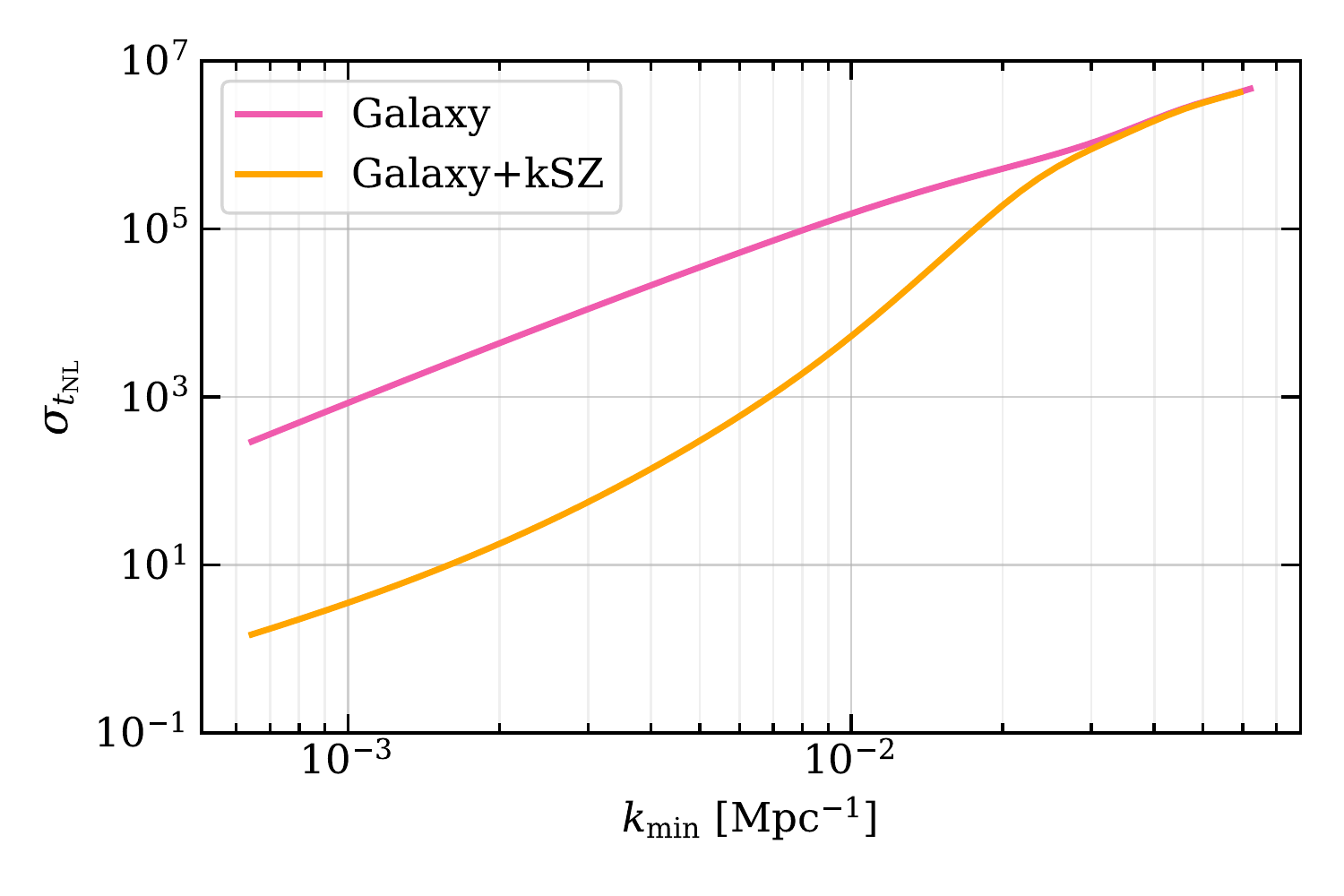}%
    \caption{\textit{Left:} $\sigma_{f_{\rm NL}}$ as a function of $k_\text{min}$ for baseline 1. \textit{Right:} $\sigma_{\tau_{\rm NL}}$ as a function of $k_\text{min}$ for baseline 1. The lower bound for $k$, on the left of the plots, is defined by the survey volume $V$. When the two data sets are cross-correlated, the error in $f_{\rm NL}$ and $\tau_{\rm NL}$ drastically decreases below $k \approx 2\times 10^{-2} \ \rm{Mpc}^{-1}$ (in comparison to the `Galaxy' case), owing to the low noise in velocity reconstruction on large scales.}%
    \label{fig:error_dep_k_min}%
\end{figure*}

To highlight the scales that contribute most to the signal, we plot both $\sigma_{f_{\rm NL}}$ and $\sigma_{\tau_{\rm NL}}$ as a function of the smallest measurable Fourier mode for our galaxy survey. This corresponds to varying the largest recoverable $k$-mode from the survey volume $V$, directly impacting the value of $k_\text{min}$ as it appears in Eq.~\eqref{eq:FisherIntegral}. These plots are displayed in Fig. \ref{fig:error_dep_k_min}. In both cases, the effects of sample variance cancellation become evident below $k \approx 2\times10^{-2} \ \text{Mpc}^{-1}$. This behaviour can be explained by comparing the model for $P_{vv}(k)$ [Eq.~\eqref{eq: Pvv model}] to the assumed model for $N_{vv}$[derived from Eq.~\eqref{eq:NvvFullExpression}]. These models indicate that the signal-to-noise ratio (SNR) of our velocity reconstruction is inversely proportional to $k^{2}$. Therefore, on small scales, we are only dependent the signal from $P_{hh}(k)$ to constrain both $f_{\rm NL}$ and $\tau_{\rm NL}$, causing the errors to coincide across the estimates from the single (galaxy) data-set and the cross-correlated (galaxy-kSZ) data-sets. In contrast, the higher SNR on larger scales allows us to constrain the non-Gaussian parameters using the models for both $P_{mh}(k)$ and $P_{hh}(k)$. Therefore, on these larger scales, the effects of sample variance cancellation are on full display. 

\begin{figure*}
    \centering
    \includegraphics[width=\columnwidth]{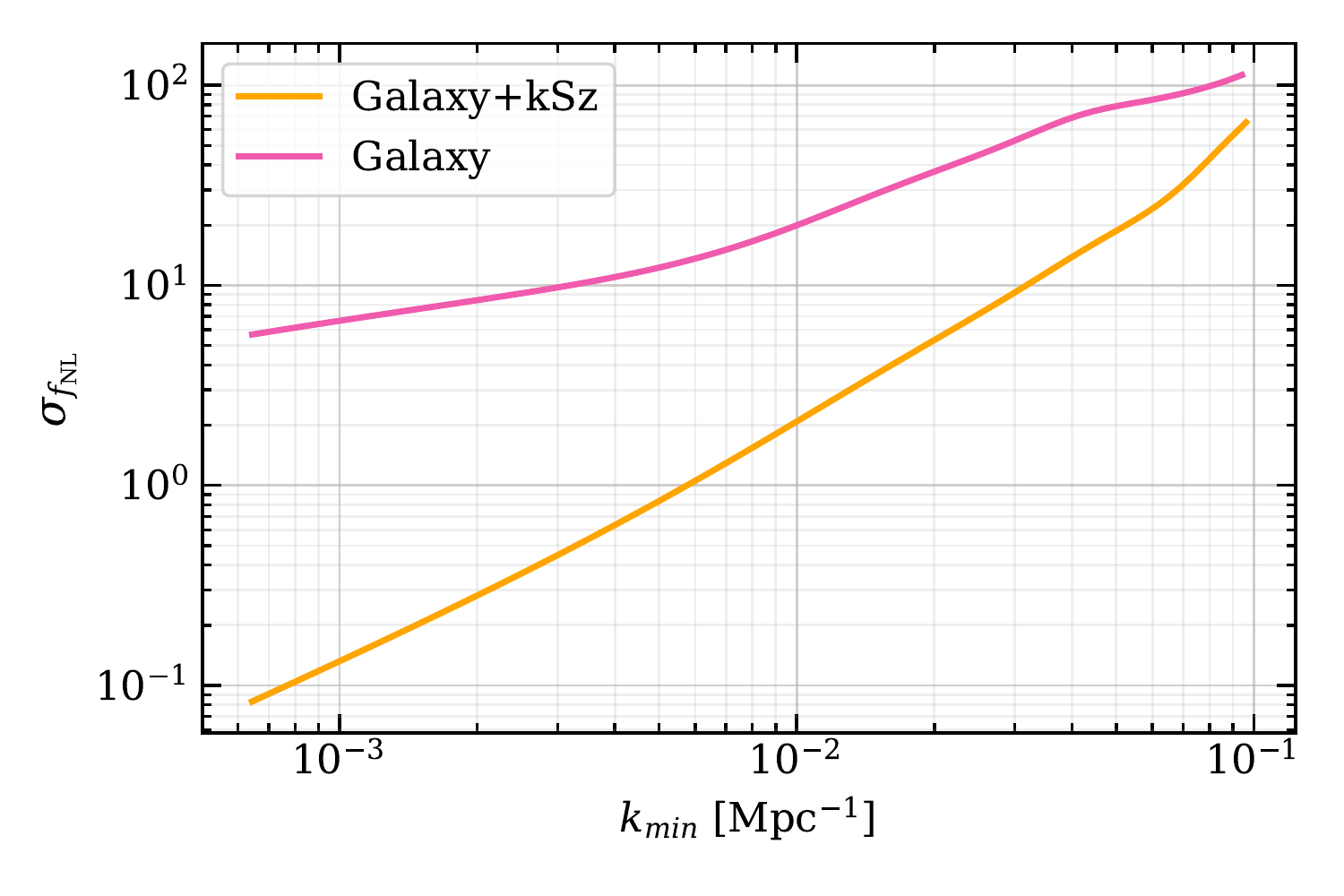}%
    \centering
    \includegraphics[width=\columnwidth]{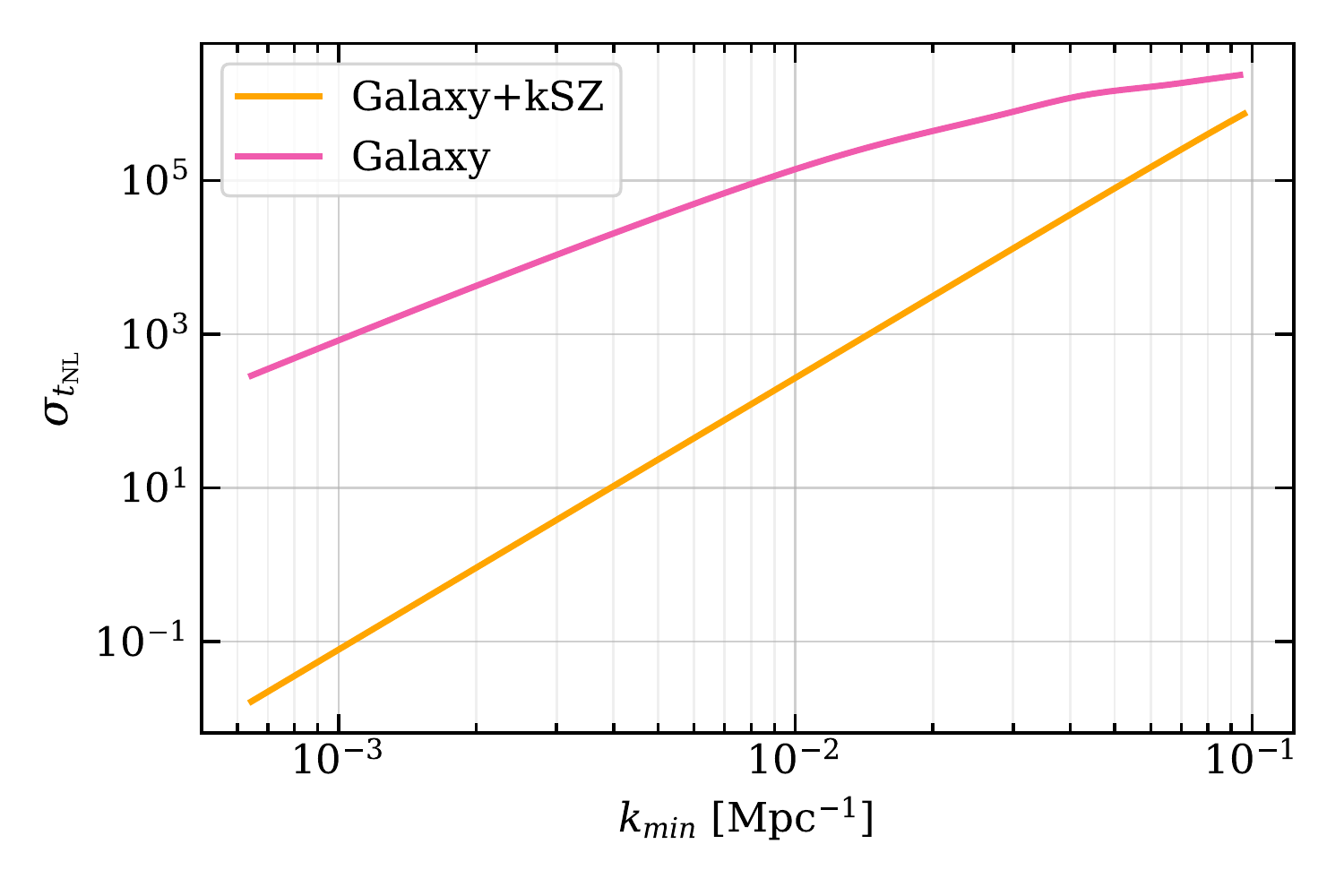}%
    \caption{\textit{Left:} $\sigma_{f_{\rm NL}}$ as a function of $k_\text{min}$ for baseline 1. \textit{Right:} $\sigma_{\tau_{\rm NL}}$ as a function of $k_\text{min}$ for baseline 1. For both the above sets of data we assume that the halo shot noise is zero and there are no photo-$z$ errors. The lower bound for $k$, at the left end of the plots, is defined by the survey volume $V$. The behaviour in this unrealistic case matches the results in Fig. \ref{fig:error_dep_k_min}, where once again we see a sharp decrease in the error in $f_{\rm NL}$ and $\tau_{\rm NL}$ when the two data sets are cross-correlated due to high SNR in the velocity reconstruction on larger scales.}%
    \label{fig:error_dep_k_min_noNoise}%
\end{figure*}

To explore, more carefully, the information contained in the signal across both cases on large scales, we also display the dependence of $\sigma_{f_{\rm NL}}$ and $\sigma_{\tau_{\rm NL}}$ on $k_\text{min}$ when there is no contamination in our signal coming from shot noise, photo-$z$ errors, and CMB instrument noise. These plots are displayed in Fig. \ref{fig:error_dep_k_min_noNoise}. 

At large values of $k$, we no longer see the `Galaxy' curve coincide with the `Galaxy+kSZ' results. This is a consequence of the much lower velocity reconstruction noise in the absence of shot noise. However, it is clear that an extension of the curves to smaller scales would reveal behaviour similar to the curves in Fig. \ref{fig:error_dep_k_min}, as a result of cosmic variance. The behaviour on larger scales is a lot more noteworthy for this set-up. In the case of $\sigma_{f_{\rm NL}}$, we see an inflection point at $k \approx 3 \times 10^{-3}\ {\rm Mpc}^{-1}$, after which the slope of the curve gets closer to zero. In contrast, when the galaxy survey data is combined with velocity reconstruction data, the forecast on $\sigma_{f_{\rm NL}}$ decreases steadily below $k\approx10^{-3}\ \text{Mpc}^{-1}$. This indicates that the large-scale modes contain a significant amount of data that allows us to better constrain $f_{\rm NL}$ with the cross-correlated data-sets. In contrast, the `Galaxy' estimate of $\sigma_{\tau_{\rm NL}}$ never experiences the inflection point seen in the corresponding $\sigma_{f_{\rm NL}}$ curve, within the range of $k_\text{min}$ plotted. However, the benefit of cross-correlating the data-sets is still evident in the relatively steeper decrease in $\sigma_{\tau_{\rm NL}}$ on larger scales. It is also important to note that when the two data sets are combined, the error in $\tau_{\rm NL}$ decreases more steeply than the error in $f_{\rm NL}$ i.e.; on the largest scales $\sigma_{\tau_{\rm NL}}$ reaches a minimum of $\sim 10^{-3}$ whereas $\sigma_{f_{\rm NL}}$ is only improved to $\sim 10^{-2}$.

\begin{figure*}
    \centering
    {\includegraphics[width=\columnwidth]{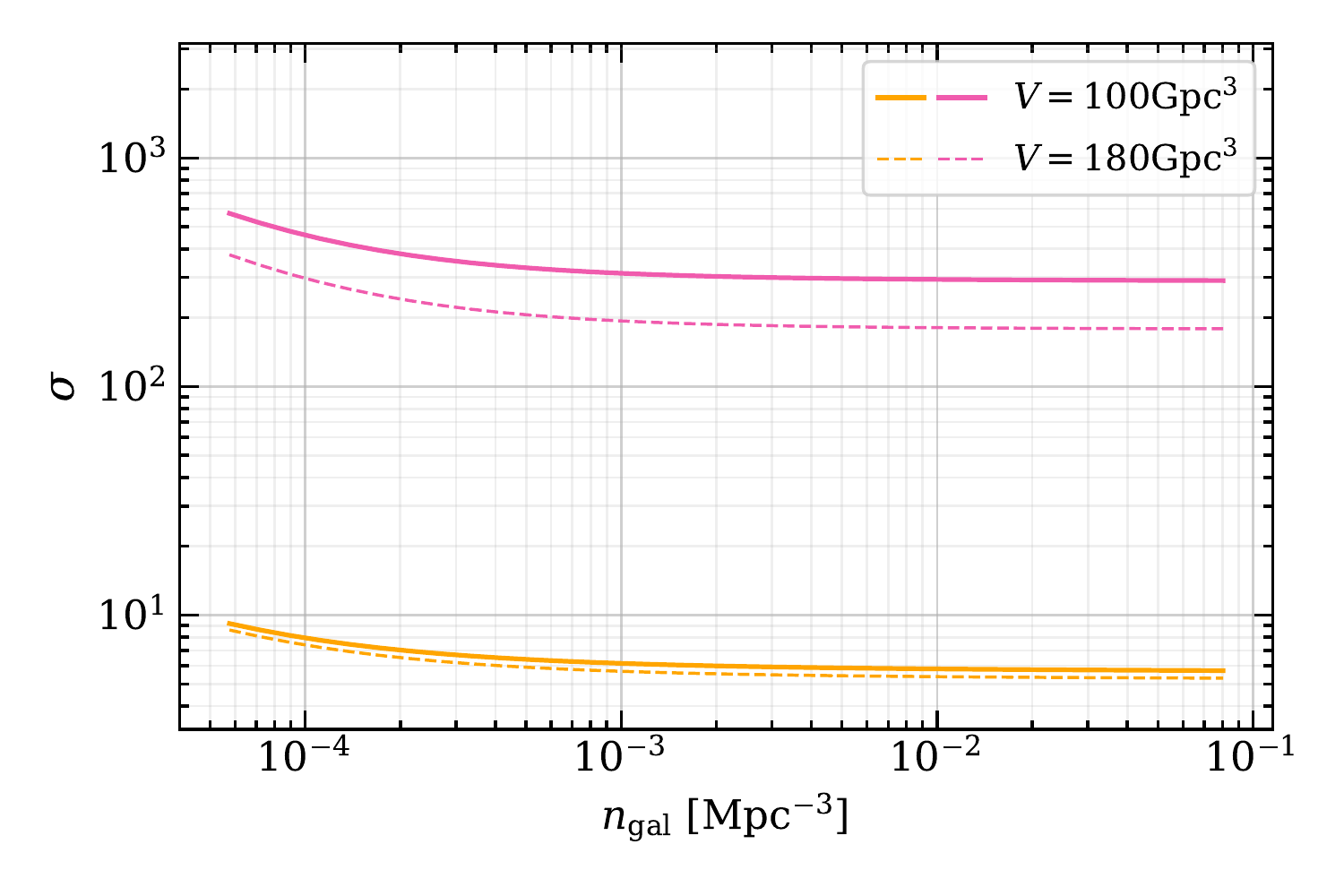} }%
    \centering
    {\includegraphics[width=\columnwidth]{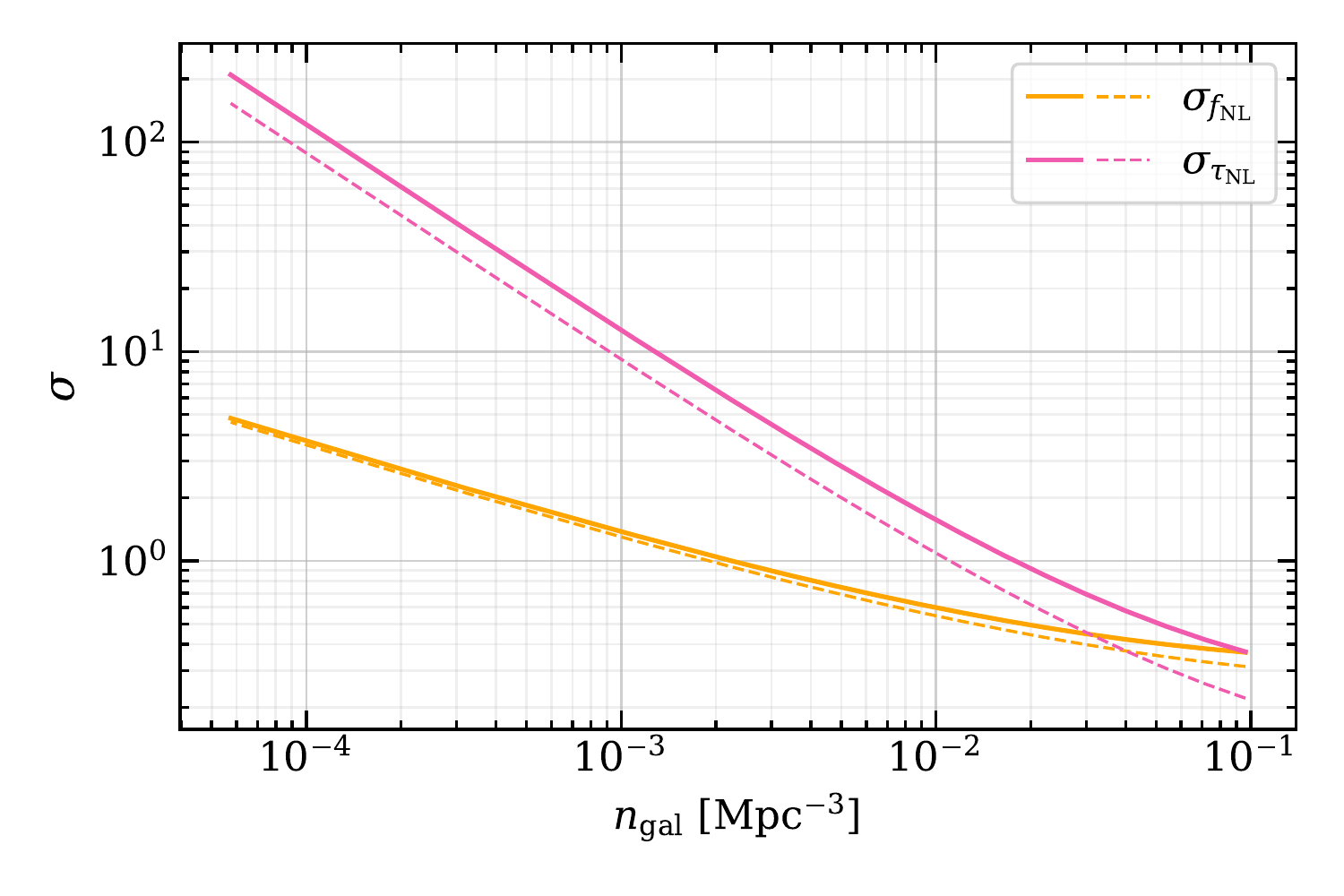} }%
    \caption{\textit{Left:} $\sigma_{f_{\rm NL}}$ and $\sigma_{\tau_{\rm NL}}$ as a function of $n_\text{gal}$, assuming that the galaxy data is used in isolation. \textit{Right:} $\sigma_{f_{\rm NL}}$ and $\sigma_{\tau_{\rm NL}}$ as a function of $n_\text{gal}$, assuming that the galaxy data is cross-correlated with the kSZ data set. The solid (dashed) lines correspond to baseline 1 with a survey volume of $100\ \rm{Gpc}^3$ ($180\ \rm{Gpc}^3$).  Galaxy density not only defines the shape of the small scale power spectra but also defines the amount of shot noise in galaxy survey data. The sharp decrease in both $\sigma_{f_{\rm NL}}$ and $\sigma_{\tau_{\rm NL}}$ for the cross-correlated data-set (right) is due to the fact that a high galaxy number density allows one to probe larger scales due to the lowered galaxy shot noise.}%
    \label{fig:error_dep_n_gal}%
\end{figure*}

The difference in the behaviour of $\sigma_{f_{\rm NL}}$ and $\sigma_{\tau_{\rm NL}}$ as a function of scale $k_\text{min}$ can be understood by analysing the non-Gaussian model for $P_{hh}(k)$ [Eq.~\eqref{eq:final_models_for_Fisher}]. The contribution of $\tau_{\rm NL}$ to this signal comes from a term that is dependent on $\alpha(k)^{-2}$, or equivalently on $k^{-4}$ [where $\alpha(k)$ is defined in Eq.~\eqref{eq:alpha_def}]. In contrast, the contribution of $f_{\rm NL}$ to this signal comes from a term that scales as $k^{-2}$. We conclude that, because the $P_{hh}(k)$ signal is dominated by the $\tau_{\rm NL}$ contribution, $\tau_{\rm NL}$ is constrained much better than $f_{\rm NL}$ on larger scales. Furthermore, this difference is most evident in the `Galaxy+kSZ' case because an improved constraint on $f_{\rm NL}$ more directly translates to an improved constraint on $\tau_{\rm NL}$ due to the inclusion of the $P_{mh}$ signal.

It is precisely this dependence on the large scales that explains the behaviour of the forecasts under varying galaxy number density $n_\text{gal}$. The dependence of $\sigma_{f_{\rm NL}}$ and $\sigma_{\tau_{\rm NL}}$ on varying values of $n_\text{gal}$ has been plotted in Fig. \ref{fig:error_dep_n_gal}. The solid lines correspond to results derived based on the experimental configuration corresponding to baseline 1 in Table \ref{tab:baselineSpecs}, including shot noise and photo-z errors. The dashed lines represent results derived from the same set-up, with only the survey volume $V$ updated to $180\ \rm{Gpc}^3$. For clarity, the fiducial value of $b_h$ was held constant. 

When considering only the galaxy survey data, the results displayed on the left in Fig. \ref{fig:error_dep_n_gal} show that although the estimates improve slightly with decreasing shot noise, the cosmic variance limit is quickly reached for both our estimates of $f_{\rm NL}$ and $\tau_{\rm NL}$, irrespective of the assumed survey volume. This is because, using galaxy survey data alone forces us to constrain both $f_{\rm NL}$ and $\tau_{\rm NL}$ using the model for $P_{hh}(k)$ in isolation. Furthermore, as seen in Fig. \ref{fig:error_dep_k_min_noNoise}, the slope in the $\sigma_{f_{\rm NL}}$ error is closer to zero with inclusion of low $k$-modes. Therefore, although the lower shot noise (higher $n_\text{gal}$) allows us to probe larger and larger scales, the ability to constrain the non-Gaussian parameters eventually plateaus, as is seen in our results.

However, when the galaxy survey data is cross-correlated with the kSZ data, the lowered shot noise has a much more pronounced impact on both $\sigma_{f_{\rm NL}}$ and $\sigma_{\tau_{\rm NL}}$ (right of Fig. \ref{fig:error_dep_n_gal}). In fact, the effect of the higher $n_\text{gal}$ on $\sigma_{\tau_{\rm NL}}$ is much steeper, with the results indicating that at high enough galaxy number density one can constrain $\tau_{\rm NL}$ better than $f_{\rm NL}$. This is because a higher number density allows for the use of more signal from large-scale $k$-modes. When cross-correlating the two data sets, the inclusion of these modes allows for a steady improvement in the ability to measure $f_{\rm NL}$ in the absence of noise, as shown in Fig. \ref{fig:error_dep_k_min_noNoise}. This improved constraint on $f_{\rm NL}$ translates into a better measurement of $\tau_{\rm NL}$. This, combined with the difference in the contribution of each of these terms to the $P_{hh}$ signal, allows for a tighter constraint on $\tau_{\rm NL}$ than $f_{\rm NL}$ when larger scales can be included as a result of lower shot noise. 

The threshold value of $n_{\rm{gal}}$ required to measure $\tau_{\rm{NL}}$ with a higher sensitivity than $f_{\rm{NL}}$ is dependent on the survey volume. The results from the two different survey volumes, presented on the right of Fig. \ref{fig:error_dep_n_gal}, indicate that the two uncertainty curves for $f_{\rm{NL}}$ and $\tau_{\rm{NL}}$ intersect at lower values of $n_{\rm{gal}}$ for larger survey volumes, as expected. While a survey with $V = 100\ \rm{Gpc}^3$ (under the baseline 1 configuration) can only achieve $\sigma_{\tau_{\rm{NL}}} \lesssim \sigma_{f_{\rm{NL}}}$ with $n_{\rm{gal}} \approx 10^{-1}\ \rm{Mpc}^{-3}$, increasing the survey volume to $V = 180\ \rm{Gpc}^3$ allows for the sensitivities to intersect at a more achievable $n_{\rm{gal}} \approx 3\times 10^{-2}\ \rm{Mpc}^{-3}$.

To establish the dependence of these errors on CMB instrumental noise [Eq.~\eqref{eq:CMBNoise}], we also calculated the values of $\sigma_{f_{\rm NL}}$ and $\sigma_{\tau_{\rm NL}}$ under varying values of sensitivity $s$ and resolution $\theta_{\rm FWHM}$, independently. In both these calculations we assumed the baseline 1 configuration for all other parameters. In our results we find that varying the sensitivity from $0.25\ \mu \rm{K}$-arcmin to $14 \ \mu \rm{K}$-arcmin approximately increases our error in $f_{\rm NL}$ by 3.1x and $\tau_{\rm NL}$ by 6.3x. In contrast, the dependence of the errors on the CMB telescope resolution is a more pronounced. When the resolution is varied from 0.5-10 arcmin, $\sigma_{f_{\rm NL}}$ steadily increases by a factor of 9.8. Similarly, the forecasted error in the estimation of $\tau_{\rm NL}$ increases steeply by a factor of 100 for the same variation in CMB telescope resolution.

Finally, the values of $\sigma_{f_{\rm NL}}$ and $\sigma_{\tau_{\rm NL}}$ were also calculated for varying values of photo-$z$ error $\sigma_z$. The value of $\sigma_z$ was varied from 0.0 to 1.0, which approximately resulted an increase in  $\sigma_{f_{\rm NL}}$ by a factor of 3.2 and an increase in $\sigma_{\tau_{\rm NL}}$ by a factor of 2.5. This minimal effect of varying $\sigma_z$ is explained by the scale dependence of the $f_{\rm NL}$ and $\tau_{\rm NL}$ terms that makes the constraints most dependent on the largest scales measured. 

\section{Conclusions}
kSZ tomography is a powerful probe of the large-scale matter distribution that will be accessible with the next generation CMB and large-scale structure surveys. Cross-correlating this data set, with galaxy distribution data from upcoming large-scale structure surveys, such as the VRO survey and DESI, leads to sample variance cancellation in the measurement of galaxy bias and other quantities. In this paper, we have calculated the sensitivity with which both $f_{\rm NL}$ and $\tau_{\rm NL}$ can be measured using this method of cross-correlation. We also display the improvement coming from the addition of the kSZ data set and identify the experimental factors that most prominently contribute to better sensitivity in our measurements. 

The statistical power of this method is most evident at large scales ($k < 10^{-2}\ \text{Mpc}^{-1}$), arising from the low noise in the velocity reconstruction from kSZ data. For a cross-correlation between VRO survey data and CMB-S4, we find that one can reach $\sigma_{f_{\rm NL}} \approx 0.59$ and $\sigma_{\tau_{\rm NL}} \approx 1.5$ and improvement factors of 10 and 195, respectively, in comparison to estimates that use VRO data alone, without internal sample variance cancellation. Similarly, for the combination of DESI and a SO like survey we calculate $\sigma_{f_{\rm NL}} \approx 3.1$ and $\sigma_{\tau_{\rm NL}} \approx 69$, with a corresponding improvement factors of 2 and 5, respectively.
This forecast includes marginalization over all relevant parameters and realistic photo-$z$ errors as well as redshift space distortions. In our analysis of the experimental parameters which most heavily influence our sensitivity to measuring the scale-dependent bias, we find that the best results are achieved when the galaxy survey data is obtained from a large survey volume, with well measured large-scale modes, in combination with a high galaxy number density count. Furthermore, we expect that binning the galaxy survey data by mass, population, or redshift, to achieve internal sample variance cancellation, will further improve sensitivity to both $\tau_{\rm{NL}}$ and $f_{\rm{NL}}$, following the analysis in \cite{Munchmeyer:2018eey, Ferraro:2014jba}.

In our work we have used a simplified 3-dimensional box geometry to illustrate the properties of the method and highlight the potential to measure signatures of non-Gaussianity using kSZ tomography data. We assume a fixed, functional form for $\beta_f$ [Eq.~\eqref{eq: beta_f}] to explicitly display forecasts on $f_{\rm NL}$ and $\tau_{\rm NL}$ alone. However, the dependence of $\beta_f$ on $b_h$ may require further simulation-based analysis for a non-Gaussian universe (see Ref. \citep{Barreira:2022sey}). Although there are other modelling assumptions intrinsic to our calculation of velocity reconstruction noise, such as the assumed distribution of electron gas within halos, we expect a marginalisation over these model parameters to have a minimal impact on our sensitivity to $f_{\rm NL}$ and $\tau_{\rm NL}$. In these forecasts we also account for optical depth degeneracy via an added parameter in our information matrix; however, we expect that the measurements of electron profiles from fast-radio burst searches~\cite{Madhavacheril:2019buy}, as well as cross-correlation between radial and transverse velocities (latter reconstructed from so called `moving-lens' tomography~\cite{Hotinli:2018yyc,Hotinli:2020ntd,Hotinli:2021hih}) can potentially mitigate this bias in the near future. Although the inclusion of GR effects could lead to degeneracies with the existing $f_{\rm NL}$ and $\tau_{\rm NL}$ parameters (explored for the $f_{\rm NL}$ case in Refs. \citep{Maartens:2020jzf, Raccanelli:2015vla}), we expect the effects of these degeneracies to be minimised by redshift binning or the consideration of multiple populations of halos.

We find that our forecasts compare well with other attempts at constraining local non-Gaussianity under sample variance cancellation, using different tools for cross-correlation. Our constraints on both $f_{\rm NL}$ and $\tau_{\rm NL}$ are slightly better than (within a factor of $\sim 2$ and $\sim 3$, respectively) the forecasts presented in Ref. \cite{Ferraro:2014jba}, where sample variance cancellation was achieved by considering multiple populations of halos, assuming an LSST-type survey. Moreover, cross-correlation of reconstructed CMB lensing potential and galaxy clustering can also probe local type of non-Gaussianities, as showin in Ref. \cite{Schmittfull:2017ffw}. Their forecast on $f_{\rm NL}$, considering the survey combination of LSST and CMB-S4, with redshift binning, is comparable to the value presented in this paper. However, such forecasts would be sensitive to lensing reconstruction biases, which are likely more detrimental \cite{Fabbian:2019tik} than similar biases in kSZ tomography \cite{Cayuso:2021ljq}. Including the CMB-lensing or moving-lens tomography \cite{Hotinli:2018yyc,Hotinli:2020ntd,Hotinli:2021hih} for additional sample variance cancellation from transverse modes could lead to some improvement, the analysis of which is left to future work. Ultimately, our forecasts in this paper indicate that kSZ tomography is a prominent tool for cross-correlation physics, allowing for impressive constraints on the PNG parameters in the curvaton scenario.

\begin{acknowledgments}
We are thankful to Mesut \c{C}al{\i}\c{s}kan for valuable discussions and help with editing the presentation of this work. SCH thanks Kendrick Smith and Mat Madhavacheril for useful conversations. GSP was supported by the National Science Foundation Graduate Research Fellowship under Grant No.\ DGE1746891.  SCH was supported by a Johns Hopkins Horizons Fellowship. This work was also supported by NSF Grant No.\ 1818899 and the Simons Foundation.
\end{acknowledgments}

\appendix

\section{Halo Model}
\label{appendix:HaloModel}
For the forecasts in this paper, we use the halo model to calculate the non-linear power spectra involving electron and galaxy fields. These power spectra are used to calculate the noise in our velocity reconstruction from the measured kSZ anisotropies. In this section, we present a short overview of this modelling methodology and present our modelling assumptions. 

The halo model is dependent on the fundamental assumption that all the dark and baryonic matter is bound in halos of varying masses and density profiles. The correlation function for the matter-density or galaxy-density fluctuations then receives two contributions, one which accounts for the clustering of distinct halos (``two-halo" term) and another which accounts for the clustering within each individual halo (``one-halo" term). A review of this model can be found in Ref. \cite{Cooray:2002dia}.

\subsection{Dark Matter}
Although the non-linear power spectrum of dark matter clustering is not directly used in the velocity reconstruction estimates, we make assumptions on the clustering of these halos that define the form of the electron and galaxy power spectra. These specifics are described below.

Given the linear matter power spectrum $P_{mm}(k)$, and the cosmological matter density $\rho_m$ (at the redshift of consideration), the rms variance of mass within a sphere of radius $R$ that contains mass $m=4\pi \rho_m R^3/ 3$ is defined as:
\begin{equation}
    \sigma^2(m, z) = \frac{1}{2\pi^2}\int_{0}^{\infty} dk \ k^2\  P_{mm}(k,z)W^2(kR).
\end{equation}
Here, $R = R(m)$ and $W(kR)$ is the window function in Fourier space:
\begin{equation}
    W(kR)  = \frac{3[\sin(kR) - kR\cos(kR)]}{(kR)^3}.
\end{equation}
This is then used to define the halo mass function,
\begin{equation}
    n(m, z) = f(\sigma,z) \frac{\rho_{m}}{m^2}\frac{d\ln[\sigma(m,z)^{-1}]}{d\ln(m)},
\end{equation}
where $m$ is the halo mass. This quantity denotes the number density of halos per mass interval, at a specific redshift $z$. For our calculations, we assume the Tinker collapse fraction \cite{Tinker:2008ff}:
\begin{equation}
    f(\sigma, z) = A\Bigg[\Big(\frac{\sigma}{b}\Big)^{-a} + 1\Bigg]e^{-c/\sigma^2},
\end{equation}
with $A=0.186$, $a = 1.47$, $b = 2.57$, and $c=1.19$. The linear halo bias, consistent with the above collapse fraction, is assumed to be \cite{Tinker:2010my}:
\begin{equation}
    \begin{split}
        b_h(\nu) = 1 + \frac{1}{\sqrt{a}\delta_c}\Big[\sqrt{a}(a\nu^2) + \sqrt{a}b(a\nu^2)^{1-c} \\ 
        - \frac{(a\nu^2)^c}{(a\nu^2)^c + b(1-c)(1-c/2)}\Big],
    \end{split}
\end{equation}
where, in this model, $a=0.707$, $b = 0.5$, $c = 0.6$, and we have defined $\nu(m,z) = \delta_c/ \sigma(m,z)$. Note that these set of equations satisfy the consistency relation:
\begin{equation}
    \int_{-\infty}^{\infty}d\ln{m}\ m^2n(m,z)\Big(\frac{m}{\rho_m}\Big)b_h(m,z) = 1.
\end{equation}

\subsection{Galaxies}
The distribution of galaxies inside each halo is modelled according to the Halo Occupation Distribution (HOD) \cite{Berlind:2001xk}. Under this model, we assume separate distributions for central and satellite galaxies, the forms of which are determined in \cite{Leauthaud:2011rj}.

The number of central galaxies in a halo is either 0 or 1. They are always located exactly at the halo's center. The mean number of centrals in a halo of mass $m$ is fixed by the amount of stellar mass in each halo and is given by:
\begin{equation}
    \bar{N}_c(m) = \frac{1}{2} - \frac{1}{2}\text{erf}\Big[\frac{\log_{10}(m_{\star}^{\text{thresh}})\ -\ \log_{10}(m_{*}(m))}{\sqrt{2}\sigma_{\log m_*}}\Big].
\end{equation}
Here, $m_{*}(m)$ is the stellar mass in each halo of mass $m$ and is modelled according to the form provided in equation 13 of Ref. \cite{Leauthaud:2011rj}. The galaxy sample is defined by imposing a threshold stellar mass $m_{\star}^{\text{thresh}}$ of observable galaxies. This model assumes a log normal distribution for stellar mass in a fixed halo of mass $m$, with a constant redshift independent scatter $\sigma_{\log m_*}$. For our calculations we set the value of this scatter to 0.2.

The mean number of satellite galaxies in a halo of mass $m$ is given by:
\begin{equation}
    \bar{N}_s(m) = \bar{N}_c(m) \Big(\frac{m}{m_{\text{sat}}}\Big)^{\alpha_\text{sat}} \text{exp}\Big(\frac{-m_{\text{cut}}}{m}\Big).
\end{equation}
The free parameters in this model, $m_\text{sat}$, $\alpha_\text{sat}$ and $m_\text{cut}$, depend on the choice of $m_{\star}^{\text{thresh}}$. Their dependence on the threshold stellar mass is consistent with the `SIGMOD1' model in Ref. \cite{Leauthaud:2011rj} at redshift $z = 1$.

The total galaxy-galaxy power spectrum is the sum of the one halo and two halo contributions, which are defined as:
\begin{widetext}
\begin{align}
    \begin{split}
    & P_{gg}^{1h}(k,z) = \int_{-\infty}^{\infty}d\ln m \ \frac{mn(m,z)}{n_\text{gal}^2}\Big[2\langle N_c(m) N_s(m)\rangle u_c(k) u_s(k|m,z) +  \langle N_s(m)\ (N_s(m)-1)\rangle u_s(k|m, z)^2\Big],\\
    & P_{gg}^{2h}(k,z) = P_{mm}(k,z)\Bigg[\int_{-\infty}^{\infty}d\ln m\ mn(m,z)b_h(m,z)\frac{\bar{N}_c(m)u_c(k)+\bar{N}_s(m)u_s(k|m,z)}{n_\text{gal}}\Bigg]^2.
    \end{split}
    \label{eq:Pgg_1h_2h}
\end{align}
\end{widetext}
where $n_\text{gal}$ is the mean number of galaxies in the simulated survey. It is dependent on the chosen value of $m_{\star}^{\text{thresh}}$, and is defined as:
\begin{equation}
    n_\text{gal} = \int_{-\infty}^\infty d\ \ln{m} \ mn(m,z)[\bar{N}_s(m) + \bar{N}_c(m)].
\end{equation}
Furthermore, $u_c(k)$ and $u_s(k|m, z)$ represent the Fourier space distribution profiles of centrals and satellite galaxies, respectively. Since we assume that the centrals are at exact halo centers, we set $u_c = 1$. We assume that the satellite galaxies follow an NFW profile:
\begin{equation}
    \rho(r|m, z) = \frac{\rho_s}{(r/r_s) (1 + r/r_s)^2},
\end{equation}
where the scale radius $r_s$ is related to the virial radius $r_\text{vir}$ via the concentration parameter $c = r_\text{vir}/r_s$. The mass and redshift dependence in this distribution arises from the assumed model for the concentration parameter:
\begin{equation}
    c(m,z) = A \Big(\frac{m}{2\times 10^{12}\ h^{-1} M_{\odot}}\Big)^{\alpha}(1+z)^\beta,
\end{equation}
where $A = 7.85$, $\alpha = -0.081$, and $\beta = -0.71$ \cite{Smith:2018bpn}. 

Finally, the expectation values $\langle N_c(m) N_s(m)\rangle$ and $\langle N_s(m)\ (N_s(m)-1)\rangle$, appearing in Eq.~\eqref{eq:Pgg_1h_2h}, are defined to be $\bar{N}_s(m)$ and $\bar{N}_s(m)^2/ \bar{N}_c(m)$, respectively, assuming $N_c(m)$ and $N_s(m)$ are maximally correlated.

\subsection{Electrons}
\begin{figure}[t!]
\includegraphics[scale=0.49]{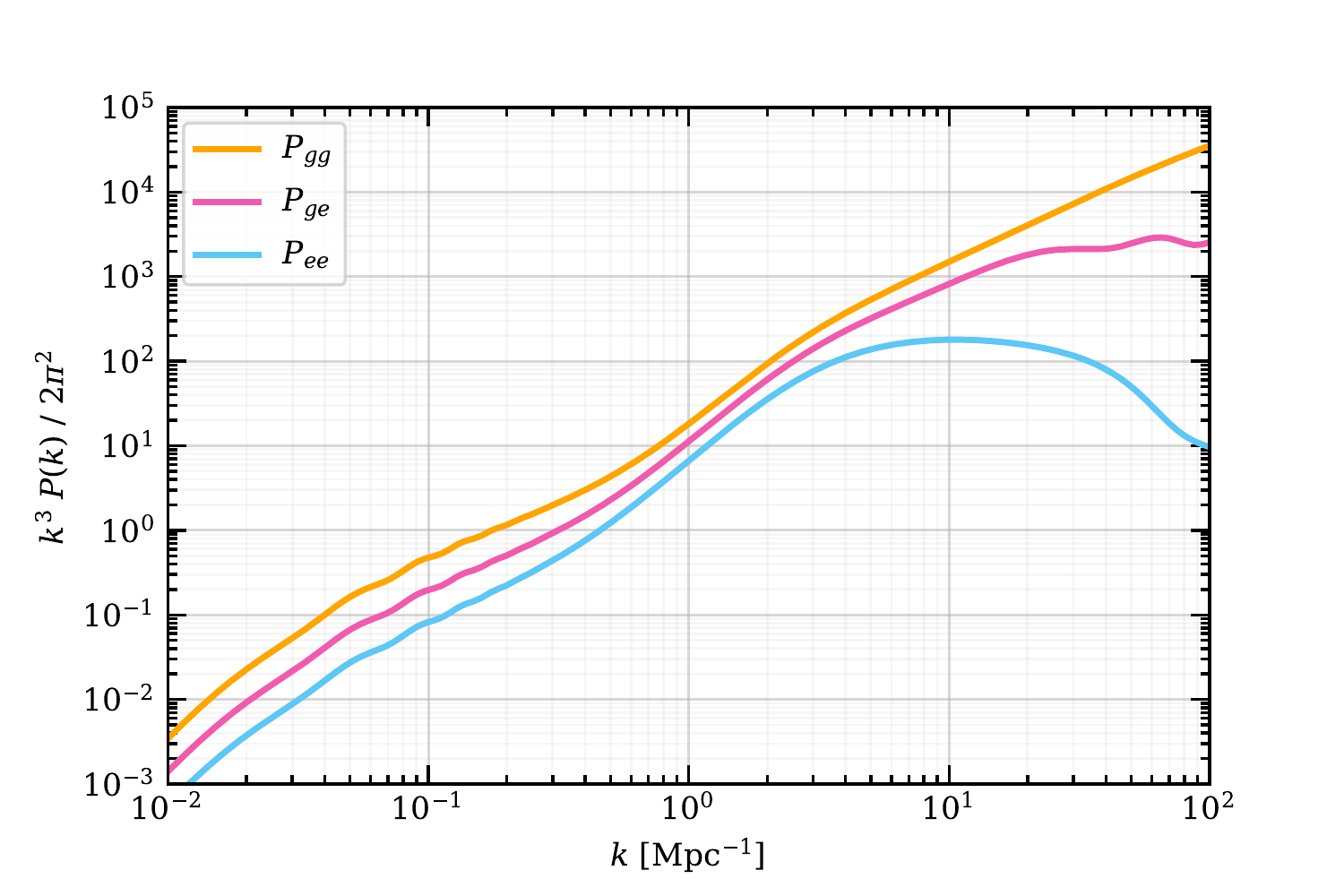} 
\caption{Auto and cross power spectra in our halo model assuming the `AGN' model of electron gas profile plotted at z=0. The shown galaxy auto-power spectrum was constructed to match the predicted galaxy number density for VRO. }
\label{fig:smallScaleSpectra}
\end{figure}

The electron distribution is modelled under the assumption that all the electron gas is bound within dark matter halos. Given this assumption, the auto-power spectrum of the electron gas is a sum of a one-halo and two-halo contribution, each of which is defined as:
\begin{widetext}
\begin{align}
    \begin{split}
        & P_{ee}^{1h}(k, z) = \int_{-\infty}^\infty d\ln{m}\ mn(m, z)\Big(\frac{m}{\rho_m}\Big)^2|u_e(k|m, z)|^2, \\
        & P_{ee}^{2h}(k,z) = P_{mm}(k, z)\Bigg[\int_{-\infty}^\infty d\ln{m}\ mn(m, z)\Big(\frac{m}{\rho_m}\Big)b_h(m, z)u_e(k| m, z)\Bigg]^2.
    \end{split}
    \label{eq:Pee_1h_2h}
\end{align}
\end{widetext}

Here, $u_e(k|m, z)$ refers to the Fourier-space distribution profile of the electron gas, which we assume to be a function of halo mass $m$ and redshift $z$ only. We use the AGN model-based fit function for the real-space mass distribution of the electron gas \cite{Battaglia:2016xbi},
\begin{equation}
    \rho_{\text{gas}} = \frac{\Omega_b}{\Omega_m}\rho_{\text{c}}(z)\bar{\rho}_0\Big(\frac{x}{x_c}\Big)^\gamma\Bigg[1 + \Big(\frac{x}{x_c}\Big)^\alpha\Bigg]^{-\frac{\beta-\gamma}{\alpha}},
\end{equation}
where we have dropped the explicit dependence of some of the above parameters on mass and redshift for ease of notation. In the above model from Ref. \cite{Battaglia:2016xbi}, $x = r/R_{200}(m, z)$ where $R_{200}$ is the radius at which the dark matter halo reaches a density of $200\rho_c(z)$. Furthermore, we have $\gamma = -0.2$ and $x_c = 0.5$. The remaining parameters $\bar{\rho}_0(m,z), \alpha(m,z)$, and $\beta(m,z)$ are fitted with a power law in halo mass and redshift:
\begin{equation}
    A = A_0^x\Bigg(\frac{m}{10^{14} M_\odot}\Bigg)^{\alpha_m^x}(1+z)^{\alpha_z^x},
\end{equation}
where the parameters for the AGN model used in this paper have been lifted from Table 2 of Ref. \cite{Battaglia:2016xbi}.

Given the auto - power spectra defined in Eq.~\eqref{eq:Pgg_1h_2h} and (\ref{eq:Pee_1h_2h}) the cross spectra can be calculated as defined in Appendix B of Ref. \cite{Smith:2018bpn}. One example set of spectra, constructed based on an $m_{\star}^{\text{thresh}}$ value that generates a galaxy number density similar to that of VRO \cite{LSSTScience:2009jmu}, has been shown in Fig.
\ref{fig:smallScaleSpectra}.

\subsection{kSZ Model}

\begin{figure}
\includegraphics[scale=0.49]{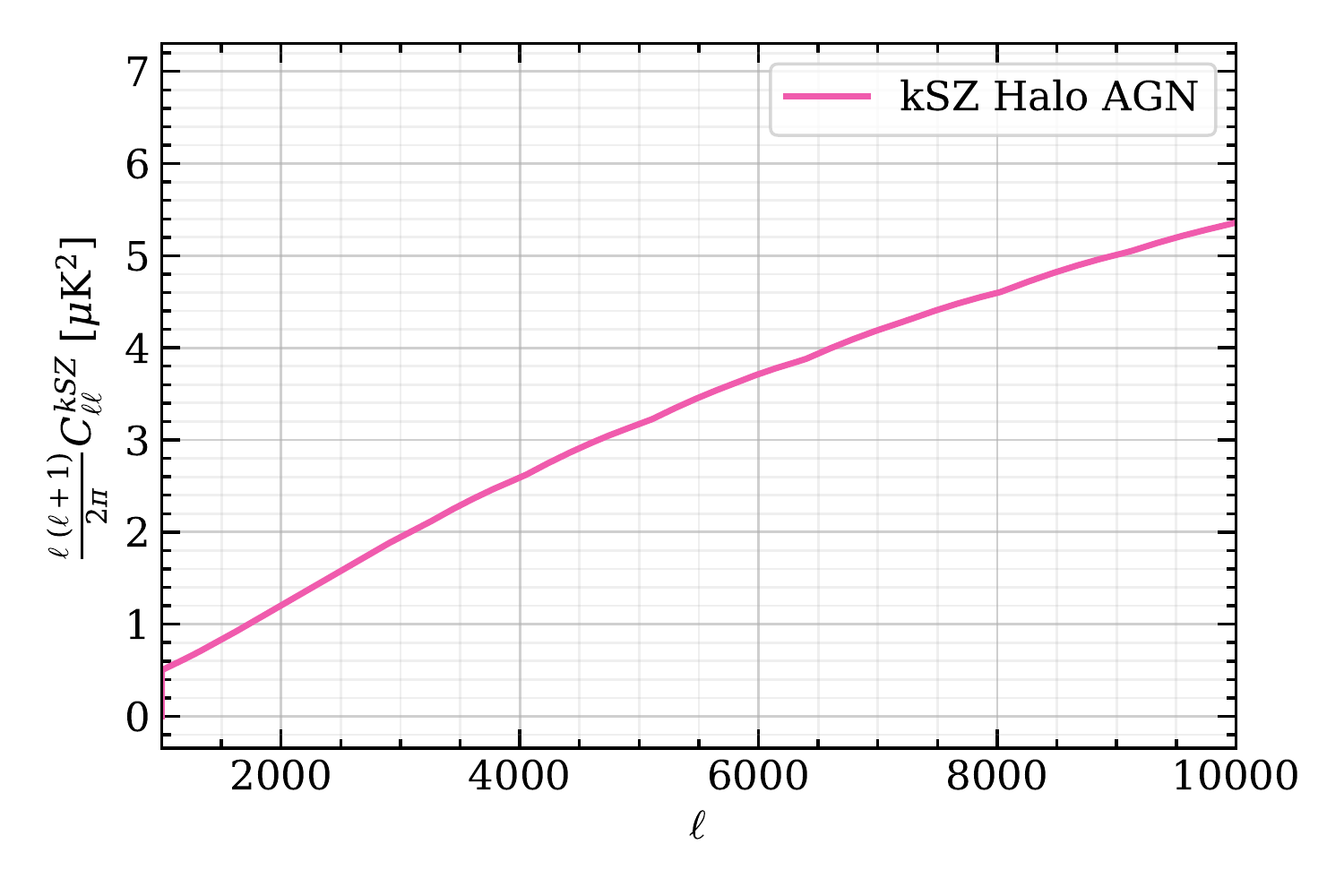} 
\caption{CMB Power spectrum for kSZ for redshifts $0 < z < 6.5$ calculated in the halo model, assuming the AGN model based electron gas density profile.}
\label{fig:lateTimekSZ}
\end{figure}

The late-time kSZ contribution to the CMB power spectrum [labelled $C_{\ell}^{\text{kSZ-late-time}}$ in Eq.~\eqref{eq:Cll_contributions}] is also modelled based on the above power spectra. The kSZ angular power spectrum at large values of $\ell$, where its contribution to the CMB spectrum is the largest, is dominated by the power spectrum of the transverse momentum field $P_{q\perp}(k)$ and is given by \cite{Vishniac:1987wm}
\begin{equation}
    C_\ell^\text{kSZ} = \frac{(\sigma_T \bar{n}_{e,0})^2}{2}\int \frac{d\chi}{\chi^2a^4} e^{-2\tau}P_{q\perp}\Big(k = \frac{\ell}{\chi}, \chi\Big).
\end{equation}
We calculate the power spectrum of the transverse momentum field based on the form provided in \cite{Hu:1999vq}
\begin{eqnarray}
    P_{q\perp}(k, z) = (fHa)^2\int_{-\infty}^{\infty} \frac{d^3k'}{(2\pi)^3}P_{ee}(|\bm{k} - \bm{k'}|, z) \\ \nonumber
\frac{k(k-2k'\mu')(1-\mu'^2)}{k'^2(k^2 + k'^2 - 2kk'\mu')}.
\end{eqnarray}
A plot of the computed $C_{\ell}^{\rm kSZ}$ used in the forecasts has been presented in Fig. \ref{fig:lateTimekSZ}.


\bibliography{apssamp}

\end{document}